\documentclass[aps,twocolumn,superscriptaddress,amsmath,amssymb,citeautoscript]{revtex4}

\usepackage{graphicx}
\usepackage{dcolumn}
\usepackage{bm}
\usepackage{multirow}
\usepackage[colorlinks=true,linkcolor=blue,urlcolor=blue, citecolor=blue]{hyperref}
\usepackage{array}
\usepackage{hyperref}

\graphicspath{{./Figures/}}

\begin{document}

\renewcommand{\arraystretch}{1.2}

\title{The thermodynamic trends of intrinsic defects in primary halide perovskites: \\ A first-principles study}

\author{Haibo Xue}
\affiliation{Materials Simulation and Modelling, Department of Applied Physics, Eindhoven University of Technology, P.O. Box 513, 5600MB Eindhoven, the Netherlands.}
\affiliation{Center for Computational Energy Research, Department of Applied Physics, Eindhoven University of Technology, P.O. Box 513, 5600MB Eindhoven, the Netherlands.}

\author{Geert Brocks}
\affiliation{Materials Simulation and Modelling, Department of Applied Physics, Eindhoven University of Technology, P.O. Box 513, 5600MB Eindhoven, the Netherlands.}
\affiliation{Center for Computational Energy Research, Department of Applied Physics, Eindhoven University of Technology, P.O. Box 513, 5600MB Eindhoven, the Netherlands.}
\affiliation{Computational Materials Science, Faculty of Science and Technology and MESA+ Institute for Nanotechnology, University of Twente, P.O. Box 217, 7500AE Enschede, the Netherlands.}

\author{Shuxia Tao}
\email{s.x.tao@tue.nl}
\affiliation{Materials Simulation and Modelling, Department of Applied Physics, Eindhoven University of Technology, P.O. Box 513, 5600MB Eindhoven, the Netherlands.}
\affiliation{Center for Computational Energy Research, Department of Applied Physics, Eindhoven University of Technology, P.O. Box 513, 5600MB Eindhoven, the Netherlands.}

\begin{abstract}
Defects in halide perovskites play an essential role in determining the efficiency and stability of the resulting optoelectronic devices. Here, we present a systematic study of intrinsic point defects in six primary metal halide perovskites, MAPbI$_3$, MAPbBr$_3$, MAPbCl$_3$, FAPbI$_3$, CsPbI$_3$ and MASnI$_3$, using density functional theory calculations with the SCAN+rVV10 functional. We analyse the impact of changing anions and cations on the defect formation energies and the charge state transitions levels and identify the physical origins underlying the observed trends. Dominant defects in the lead-iodide compounds are the A$^+$ cation interstitials (A = Cs, MA, FA), charge-compensated by I$^-$ interstitials or lead $({2-})$ vacancies. In the lead-bromide and -chloride compounds, halide vacancies become relatively more prominent, and for MAPbBr$_3$, the Pb$^{2+}$ interstitial also becomes important. The trends can be explained in terms of the changes in electrostatic interactions and chemical bonding upon replacing cations and anions. Defect physics in MASnI$_3$ is strongly dominated by tin $({2-})$ vacancies, promoted by the easy oxidation of the tin perovskite. Intrinsically, all compounds are mildly p-doped, except for MASnI$_3$, which is strongly p-doped. All acceptor levels created by defects in the six perovskites are shallow. Some defects, halide vacancies and Pb or Sn interstitials in particular, create deep donor traps. Although these traps might hamper the electronic behavior of MAPbBr$_3$ and MAPbCl$_3$, in iodine-based perovskites their equilibrium concentrations are too small to affect the materials' properties.
\end{abstract}

\maketitle

\section{Introduction}\label{section:Introduction}

Metal halide perovskites are emerging semiconducting materials for optoelectronic applications, with a general composition AMX$_3$, where A = MA (methylammonium), FA (formamidinium), or Cs; M = Pb or Sn; X = I, Br, or Cl \cite{Jena2019, Saparov2016, Stoumpos2013}. Applications of these materials in photovoltaics have drawn particular attention, as the photoelectric conversion efficiency of metal halide perovskite solar cells (PSCs) has grown spectacularly from 3.8\% in 2009 \cite{Kojima2009} to 25.5\% in 2020 \cite{NREL2021}. On the downside, the materials are hampered by long-term instabilities \cite{Juarez-Perez2019, Xu2018, Song2018, Juarez-Perez2018, Chun-RenKe2017, Latini2017, Juarez-Perez2016, Jung2016, Dualeh2014}, leading to severe loss of solar cell efficiency \cite{Doherty2020, Ono2020, Heo2017, Blancon2016, Leijtens2016, Shi2015, DeQuilettes2015}. Because of the soft nature of the perovskites crystals, formation of defects is unavoidable during the materials' growth and the device operation, and the materials' degradation is likely triggered by these defects \cite{Chen2020, Yuan2016, Ball2016, Chen2016a, Adinolfi2016, Blancon2016, Leijtens2016, Shi2015, DeQuilettes2015, Samiee2014a}.

One of the widely employed strategies for defect management is composition engineering \cite{Ono2020, Chen2020}. PSCs can achieve a higher efficiency and be more stable through the use of a mixture of cations and/or anions within the perovskite structure\cite{Qin2020, Li2020, Zhou2019, Yang2017, Saliba2016}. By finding the right mix, the formation of defects can be suppressed, and the number of charge carrier recombination centers can be reduced \cite{Chen2020, Ono2020, Zhou2019, Jena2019}. However, composition engineering often results in several inter-twinned changes that also impact the structure of the final perovskite films, leaving its exact function in defect control ambiguous. To achieve a better fundamental understanding, characterizing the precise nature of the defects becomes an important task.

Due to the limitations of experimental techniques in identifying different defect types, or in characterizing their microscopic structure \cite{Ono2020, Hieulle2018, Shih2017, Adinolfi2016, Ohmann2015}, the study of defects has greatly benefited from computational modelling based on density functional theory (DFT), in particular for the elementary point defects \cite{Yin2014, Meggiolaro2018review, Daniele2018iodine, Buin2015, Shi2015a, Meggiolaro2020a, Shi2017, Liu2018, Huang2018, Kang2017, Xu2014, Liang2021, Du2015, Michael2014}. 
Many studies have focused on a single perovskite material AMX$_3$ \cite{Yin2014, Meggiolaro2018review, Daniele2018iodine, Shi2015a, Liu2018, Huang2018, Kang2017, Xu2014, Liang2021}, or on a set of materials where either the anion X or the cation A or M is varied \cite{Shi2017, Meggiolaro2020a, Buin2015}.
Such studies have been performed using a diversity of structural models and computational methods, specifically DFT functionals, which makes it challenging to compare the results of different studies quantitatively. 

Defect levels and thermodynamics can be sensitive to the DFT functional used, for instance \cite{Xue2021, Meggiolaro2018review}. In a previous study on point defects in MAPbI$_3$ we have observed that, in order to obtain accurate results for defect properties, it is important to use the same functional for both geometry and electronic structure optimizations in a self-consistent way \cite{Xue2021}. Moreover, in the same study we have concluded that including van der Waals (vdW) terms in the functional is essential. These long-range, non-bonding, interactions have an opposite effect on the formation energies of the two quintessential types of point defects, vacancies and interstitials. 

In this work, we calculate and compare defect properties of six primary metal halide perovskites, MAPbI$_3$, MAPbBr$_3$, MAPbCl$_3$, FAPbI$_3$, CsPbI$_3$ and MASnI$_3$ using a single computational technique, in order to systematically study the effect of different compositions on the defect thermodynamics. The changes in the defect formation energies and the charge state transition levels, upon variation of the X anion, A cation and M cation are compared and analyzed. The dominant defects in each perovskite are identified and their electronic nature is examined.

\section{Computational approach}

\subsection{DFT calculations} \label{sec: DFT calculations}
Density functional theory (DFT) calculations are performed with the Vienna ab initio simulation package (VASP) \cite{Kresse1993, Kresse1996, Kresse1996a}, employing the SCAN+rVV10 functional for electronic calculations and geometry optimization \cite{Peng2016}. This functional combines the strongly constrained and appropriately normed (SCAN) \cite{Sun2015} meta-generalized gradient approximation (meta-GGA) functional with the long-range van der Waals interactions from the revised Vydrova - van Voorhis non-local correlation functional (rVV10) \cite{Sabatini2013}. It has emerged in a previous study as a reliable functional for calculating defect properties of metal halide perovskites\cite{Xue2021}.

We omit spin-orbit coupling (SOC), as we have shown that SOC has little effect on the formation energies of defects\cite{Xue2021}. Our calculations use a plane wave kinetic energy cutoff of 500 eV and a $\Gamma$-point only \textbf{k}-point mesh. The energy and force convergence criteria are set to 10$^{-4}$ eV and 0.02 eV/\AA, respectively. Bond orders of covalent bonds and atomic charges are calculated using the Density Derived Electrostatic and Chemical (DDEC6) charge partitioning method, as implemented in the Chargemol code \cite{Manz2016, Manz2017, Manz2016program}. 

\subsection{Structures} \label{sec: structures}
For MA- and FA-based perovskites, we choose the tetragonal phase as a starting point, while for the Cs-based perovskites the orthorhombic phase is used \cite{Tao2019}. Each of these structures contains four AMX$_3$ formula units per unit cell. Defective structures are created starting from 2$\times$2$\times$2 supercells, which contain 32 formula units per supercell. The structures of MAPbI$_3$, CsSnI$_3$ and FAPbI$_3$ are taken from Refs. \onlinecite{walsh_2015,walsh_2016,walsh_2017}, and reoptimized with the SCAN+rVV10 functional, including reoptimizing the volume of the unit cell. 

We have noticed that, upon introduction of defects, the FA cations in FAPbI$_3$ in the whole supercell rotate, accompanied by a significant distortion of the PbI$_6$ octahedral framework. This indicates that the starting structure of FAPbI$_3$ is not the actual lowest energy state with the SCAN+rVV10 functional. No such elaborate distortions are observed in MAPbI$_3$ or CsSnI$_3$. This is possibly due to the inclusion of vdW interactions, which are relatively more important for the large FA ion than for the smaller MA or Cs ions. Note that vdW interactions are missing in the PBE functional that has been used in Ref. \onlinecite{walsh_2017} to generate the structure of FAPbI$_3$. 

For FA-based perovskites we therefore generate new structures by removing the defects again from the optimized structures, and reoptimize the atomic positions and volume. The structures with the lowest total energies are then selected for the further study of defects. A more detailed description of the procedure and analysis is described in the Supplemental Material, Figs. S1 and S2.

Structures for MAPbBr$_3$, MAPbCl$_3$, and MASnI$_3$ are created by substituting I in MAPbI$_3$ with Br/Cl or Pb with Sn. Similarly, the CsPbI$_3$ structure is created by substituting Sn in CsSnI$_3$ with Pb. These structures, including the volume of the unit cell, are then optimized with the SCAN+rVV10 functional.

An interstitial is created by adding to the supercell a cation or an anion in a specific charge state, and then optimize the atomic positions within the supercell. Likewise, a vacancy is created by removing from the supercell a cation or an anion. In order to find the most preferable site for each defect, different nonequivalent positions in each perovskite are tested using a strategy similar to that in our previous work \cite{Xue2021}. Subsequently, the structures with the lowest total energy are selected.

\subsection{Defect formation energy} \label{sec: DFE}
An important physical property typically used to characterize point defects, is the defect formation energy (DFE) \cite{Walle2004, Freysoldt2014}. The DFE determines which defects are dominantly present under thermodynamic equilibrium conditions. Defects with smaller formation energies are easier to form, and are present in higher concentrations. Such defects are then more likely to play a role in the materials' degradation processes. 

\begin{figure*}
    \includegraphics[width=1\textwidth]{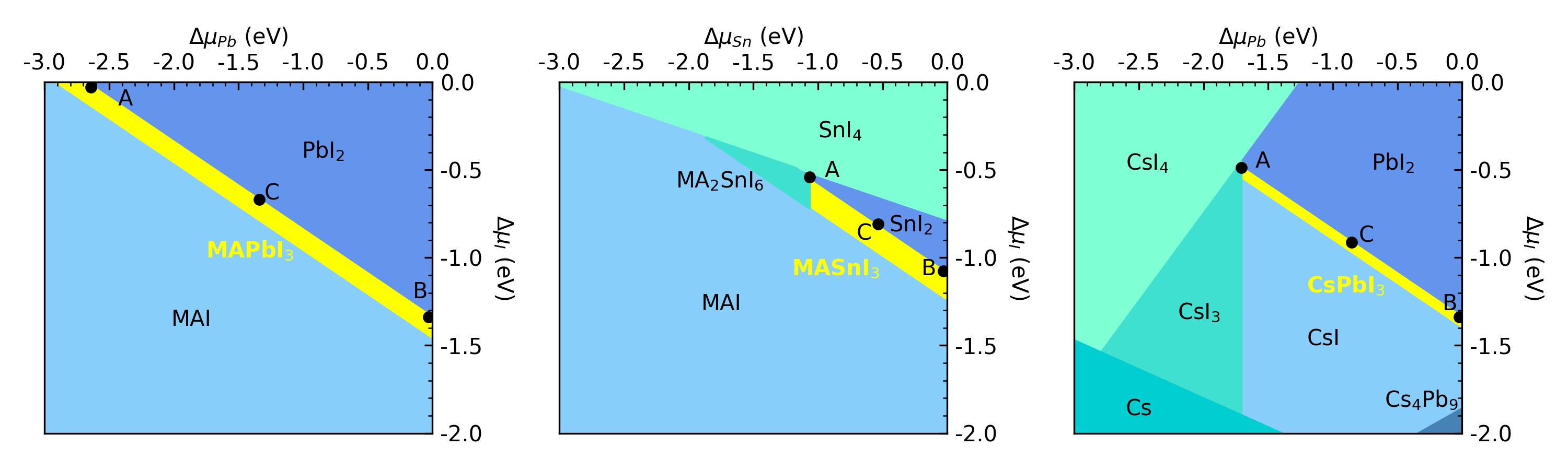}
    \caption{Calculated phase diagrams of (a) MAPbI$_3$, (b) MASnI$_3$, and (c) CsPbI$_3$. The phase diagrams of MAPbBr$_3$, MAPbCl$_3$, and FAPbI$_3$ are qualitatively similar to (a). $\mu_\mathrm{I} = \mu_\mathrm{I_2,molecule}/2$ corresponds to  $\Delta \mu_\mathrm{I} = 0$ in the figures, and $\mu_\mathrm{M} = \mu_\mathrm{M,bulk}$ defines $\Delta \mu_\mathrm{M} = 0$ for M = Pb, Sn. The points A and B define iodine-rich and iodine-poor conditions, respectively. Iodine-medium conditions (point C) are defined as halfway between points A and B.}
    \label{fig:chemicalpotentials}
\end{figure*}

The DFE $\Delta H_{f}$ is calculated from the expression \cite{Walle2004}
\begin{equation}\label{eq:DFE}
    \begin{aligned}
        \Delta H_{f}(D^q) = &E_\mathrm{tot}(D^q)-E_\mathrm{bulk} - \sum_{i} n_{i} \mu_{i} \\ 
        &+ q(E_{F}+ E_\mathrm{VBM}+ \Delta V) + E^q_\mathrm{corr},
    \end{aligned}
\end{equation}
where $D^q$ indicates the type of defect with charge $q$,  $E_\mathrm{tot}(D^q)$ and $E_\mathrm{bulk}$ are the DFT total energies of the defective and pristine supercells, respectively, and $n_i$, $\mu_i$ specify the number of atoms and chemical potential of species $i$ added to ($n_i>0$) or removed from ($n_i<0$) the pristine supercell in order to create the defect. Creating a charge $q$ requires taking electrons from or adding them to a reservoir at a fixed Fermi level, as indicated by the second line in Eq. (\ref{eq:DFE}).  

The Fermi energy is calculated as $E_{F}+ E_\mathrm{VBM}$, with $0\leq E_F \leq E_g$, the band gap, and $E_\mathrm{VBM}$ the energy of the valence band maximum. As it is difficult to determine the latter from a calculation on a defective cell, one establishes $E_\mathrm{VBM}$ in the pristine cell, shifted by $\Delta V$, which is calculated by lining up the core level on one same atom in the pristine and the neutral defective cell that is far from the defect \cite{Walle2004, Komsa2012}.

$E^q_\mathrm{corr}$ corrects for the electrostatic interaction between a charged point defect and its periodically repeated images. As in Refs. \onlinecite{Meggiolaro2018review} and \onlinecite{Xue2021} we find that the 2$\times$2$\times$2 tetragonal supercell and the dielectric screening in AMX$_3$ perovskites are sufficiently large, such that this correction is small and can be neglected. We neglect vibrational contributions to the DFE as well as the influence of thermal expansion on it, as these are typically small in the present compounds \cite{Xue2021, Wiktor2017}.

The chemical potentials $\mu_i$ of atomic species $i$ are calculated by assuming that AMX$_3$ is stable, so using $\mu _\mathrm{A} + \mu _\mathrm{M} + 3 \mu _\mathrm{X} = \mu_\mathrm{AMX_3}$ as a constraint, where for $\mu_\mathrm{AMX_3}$ we use the DFT total energy per formula unit of the AMX$_3$ perovskite. Furthermore, we assume that the perovskite is in equilibrium with the MX$_2$ phase, so $\mu_\mathrm{M} + 2 \mu_\mathrm{X} = \mu_\mathrm{MX_2}$, with $\mu_\mathrm{MX_2}$ the DFT total energy per formula unit of $\mathrm{MX_2}$. All $\mu_i$ can now be expressed in terms of a single parameter, $\mu_\mathrm{X}$, which is constrained by $\mu_\mathrm{MX_2} - \mu_\mathrm{M,bulk} \leq 2 \mu_\mathrm{X} \leq \mu_\mathrm{X_2,molecule}$. The outer bounds define X-poor (or M-rich) or X-rich conditions, respectively, with $\mu_\mathrm{M,bulk}$ and $\mu_\mathrm{X_2,molecule}$ the DFT total energies of bulk M metal, and a X$_2$ molecule. X-poor and X-rich conditions are indicated by points B and A, respectively, in Fig. \ref{fig:chemicalpotentials}(a). 

For MASnI$_3$ and CsPbI$_3$, the allowed interval for $\mu_\mathrm{X}$ has to be narrowed down to prevent the formation of other phases, such as SnI$_4$ or $\mathrm{MA_2SnI_6}$ for MASnI$_3$ \cite{Shi2017, Meggiolaro2020a}, see Fig. \ref{fig:chemicalpotentials}(b), and CsI$_3$ or CsI$_4$ for CsPbI$_3$ \cite{Li2017a, Huang2018}, see Fig. \ref{fig:chemicalpotentials}(c). In the calculations presented in this paper we choose X-medium conditions, as indicated by points C in Fig. \ref{fig:chemicalpotentials}, unless explicitly mentioned otherwise. The exact values of the chemical potentials are given in the Supplemental Material, Table S1. It should be noted that, when studying defects under intrinsic conditions, the exact choice of $\mu_\mathrm{X}$ is not so important, because a change in $\mu_\mathrm{X}$ is exactly compensated by a change in intrinsic Fermi level $E_F^{(i)}$\cite{Xue2021}. 

The intrinsic Fermi level can be determined by the charge neutrality condition, which expresses the fact that, if no charges are injected in a material, it has to be charge neutral
\begin{equation} \label{eq:charge_neutrality}
    p - n + \sum_{D^q} q \; c(D^q) = 0,
\end{equation}
where $p$ and $n$ are the intrinsic charge densities of holes and electrons of the semiconductor material, $c(D^q)$ is the concentration of defect $D^q$, and the sum is over all types of charged defects. The concentrations can be estimated by Boltzmann statistics
\begin{equation} \label{eq:defect_concentration}
    c(D^q)=c_0(D^q)  \exp \left[- \frac{\Delta H_f (D^q)}{k_BT} \right],
\end{equation}
where $c_0(D^q)$ is the density of possible sites for the defect, $T$ is the temperature, $k_B$ is the Boltzmann constant, and $\Delta H_f (D^q)$ follows from Eq. (\ref{eq:DFE}). Obviously, $p$, $n$, and $c(D^q)$ are functions of $E_F$, so the charge neutrality condition, Eq. (\ref{eq:charge_neutrality}), serves to determine the intrinsic position of the Fermi level $E_F^{(i)}$. Note that, if $E_F^{(i)}$ is sufficiently far from the band edges, then $p$ and $n$ are small, and can be neglected.

\begin{figure*}
    \includegraphics[width=1\textwidth]{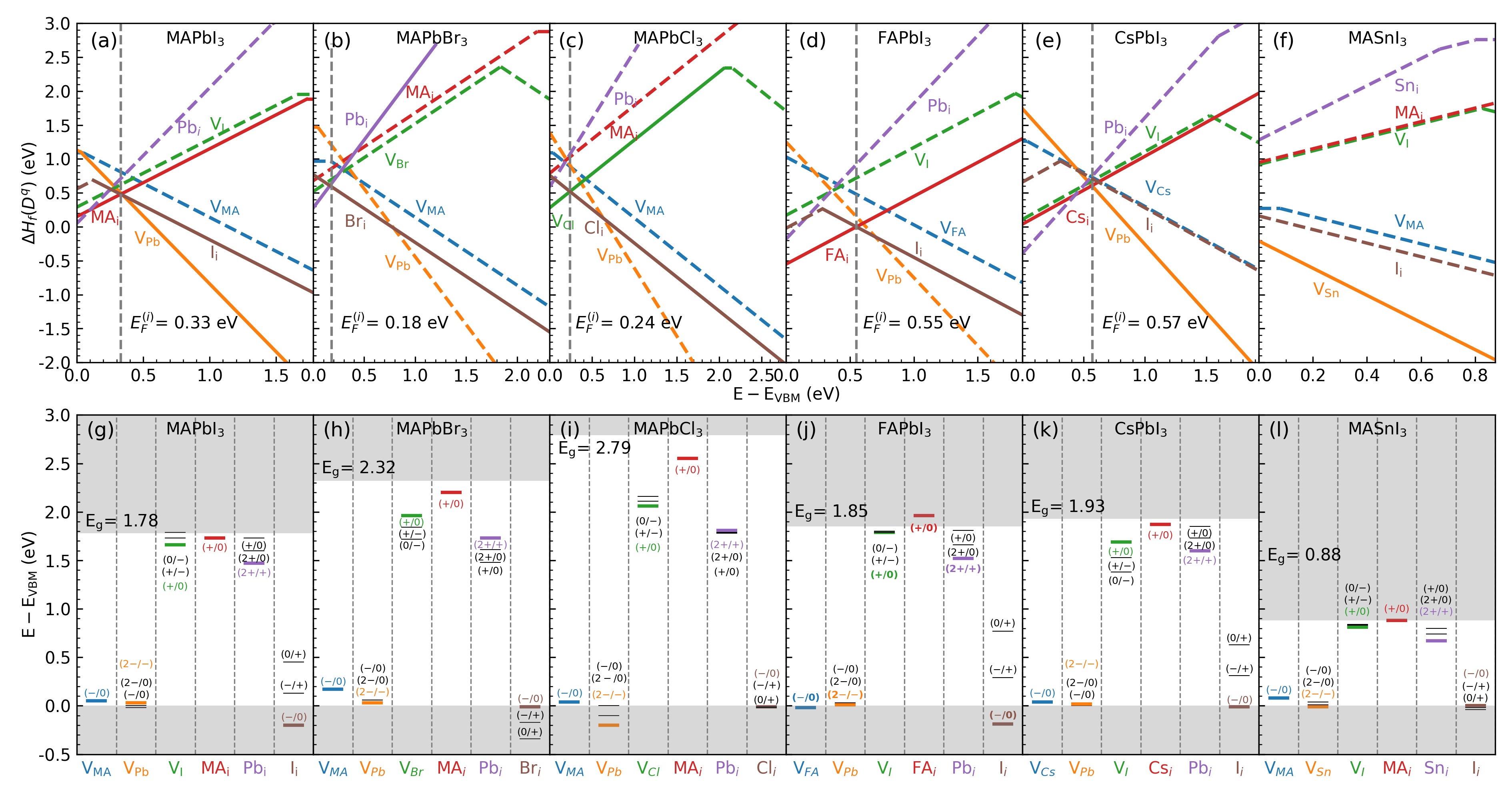}
    \caption{(a-f) Defect formation energies in six AMX$_3$ perovskites calculated at halide-medium conditions, using the SCAN+rVV10 functional; the vertical dashed lines represent the intrinsic Fermi level; the solid lines represent the energetically most favorable defects, and the dashed lines other possible defects. (g-l) Charge state transition levels; the most important ones are indicated by colored lines, representing a change of a single unit $\pm e$ starting from one stable charge state of a defect \cite{Xue2021}; the bottom and top gray areas represent the valence and conduction bands (calculated with SCAN+rVV10 without SOC), artificially aligned at the VBM.}
    \label{fig: DFEs and CSTLs}
\end{figure*}

\subsection{Charge state transition level} \label{sec: CSTL}

Under operating conditions, charges are injected in the material, shifting the positions of the (quasi) Fermi levels for electrons and holes. The charge state transition level (CSTL) $\varepsilon(q/q')$ is defined as the Fermi level position where the charge states $q$ and $q'$ of the same type of defect have equal formation energy, $\Delta H_{f}(D^q)=\Delta H_{f}(D^{q'})$. As the DFEs have a simple linear dependence on $E_F$, Eq. (\ref{eq:DFE}), this condition can be expressed as
\begin{equation} \label{eq:CSTL}
    \varepsilon(q/q')= \frac{\Delta H_{f}(D^q,E_{F}=0)-\Delta H_{f}(D^{q'},E_{F}=0)}{q'-q},
\end{equation}
where $\Delta H_{f}(D^q,E_{F}=0)$ is the DFE calculated at $E_F = 0$. 

The CSTLs influence the electronic properties of the semiconducting perovskites. If these levels are deep inside the bandgap, they are able to trap charge carriers, and act as non-radiative recombination centers. 

\section{Results and discussion}\label{section: results and discussion}

Following the strategy outlined in the previous section, we calculate the defect thermodynamics and electronic properties of the halogen vacancy V$_\mathrm{X}$, the metal vacancy V$_\mathrm{M}$, and the A-cation vacancy V$_\mathrm{A}$ in the perovskites AMX$_3$, and of their corresponding interstitials X$_i$, M$_i$, and A$_i$. All point defects are studied in the neutral, as well as in all physically reasonable charged states. We study the six elementary perovskites MAPbI$_3$, MAPbBr$_3$, MAPbCl$_3$, FAPbI$_3$, CsPbI$_3$, and MASnI$_3$, which enables us to assess the effect of halide substitution (from the first three compounds in this series), of A-cation substitution (comparing FAPbI$_3$ and CsPbI$_3$ to MAPbI$_3$), and of metal substitution (comparing MASnI$_3$ and MAPbI$_3$).

The calculated DFEs and CSTLs are shown in Fig. \ref{fig: DFEs and CSTLs}, which will be discussed in more detail below. We will focus on the energies; optimized geometries of the point defects in different charge states are visualized in the Supplemental Material, Fig. S4. The results shown in Fig. \ref{fig: DFEs and CSTLs} are calculated for halide-medium growth conditions, the points marked C in Fig. \ref{fig:chemicalpotentials}. For comparison, the DFEs of the AMX$_3$ perovskites at halide-rich and -poor conditions (points A and B in Fig. \ref{fig:chemicalpotentials}) are shown in the Supplemental Material, Figures S5-S7. 

Fig. \ref{fig: DFEs and CSTLs}(a)-(f) show the DFEs,  $\Delta H_{f}(D^q)$, of all vacancies and interstitials as a function of the position of the Fermi level, $E_F$, with respect to the position of the VBM, $E_\mathrm{VBM}$. All curves show a linear relation, as $\Delta H_{f}(D^q) \propto qE_F$, where $q$ is the charge state of the defect, Eq. \ref{eq:DFE}. For each defect and $E_F$ the curve of the most stable defect is shown as a solid line. The intrinsic Fermi level, $E_F^{(i)}$, is found by obeying the charge neutrality condition, Eq. \ref{eq:charge_neutrality}. 

The points in Fig. \ref{fig: DFEs and CSTLs}(a)-(f) where the curves discontinuously change slope are in fact crossing points between two curves belonging to different charge states of the same defect. These points mark the CSTLs, where a defect changes its charge state, Eq. \ref{eq:CSTL}. The CSTLs are visualized in Fig. \ref{fig: DFEs and CSTLs}(g)-(l) with respect to $E_\mathrm{VBM}$ and $E_\mathrm{VBM} + E_g$. Note that for ease of comparison the VBMs of different perovskites are aligned to the same position.

Using the results shown in Fig. \ref{fig: DFEs and CSTLs}, we will analyze in the following sections the effects on the defect physics of the perovskite AMX$_3$, upon varying the halide (X anion), the A cation and the metal (M cation). We use MAPbI$_3$ as a point of reference, whose calculated DFEs are shown in Fig. \ref{fig: DFEs and CSTLs}(a). Close to intrinsic conditions ($E_F^{(i)}=0.33$ eV), point defects in MAPbI$_3$ have a preference for a specific charge state, interstitials appear as anions or cations, $\mathrm{I_i}^-$, $\mathrm{MA_i}^+$, and $\mathrm{Pb_i}^{2+}$, and vacancies have a charge opposite to their anion/cation interstitial counterparts, $\mathrm{V_I}^+$, $\mathrm{V_{MA}}^-$ and $\mathrm{V_{Pb}}^{2-}$. Defects in other halide perovskites have the same pattern of charges close to intrinsic conditions, as shown in Fig. \ref{fig: DFEs and CSTLs} (b)-(f).

\subsection{Varying the X anion} \label{subsection: varying the X anion}

\begin{figure}
    \includegraphics[width=1\columnwidth]{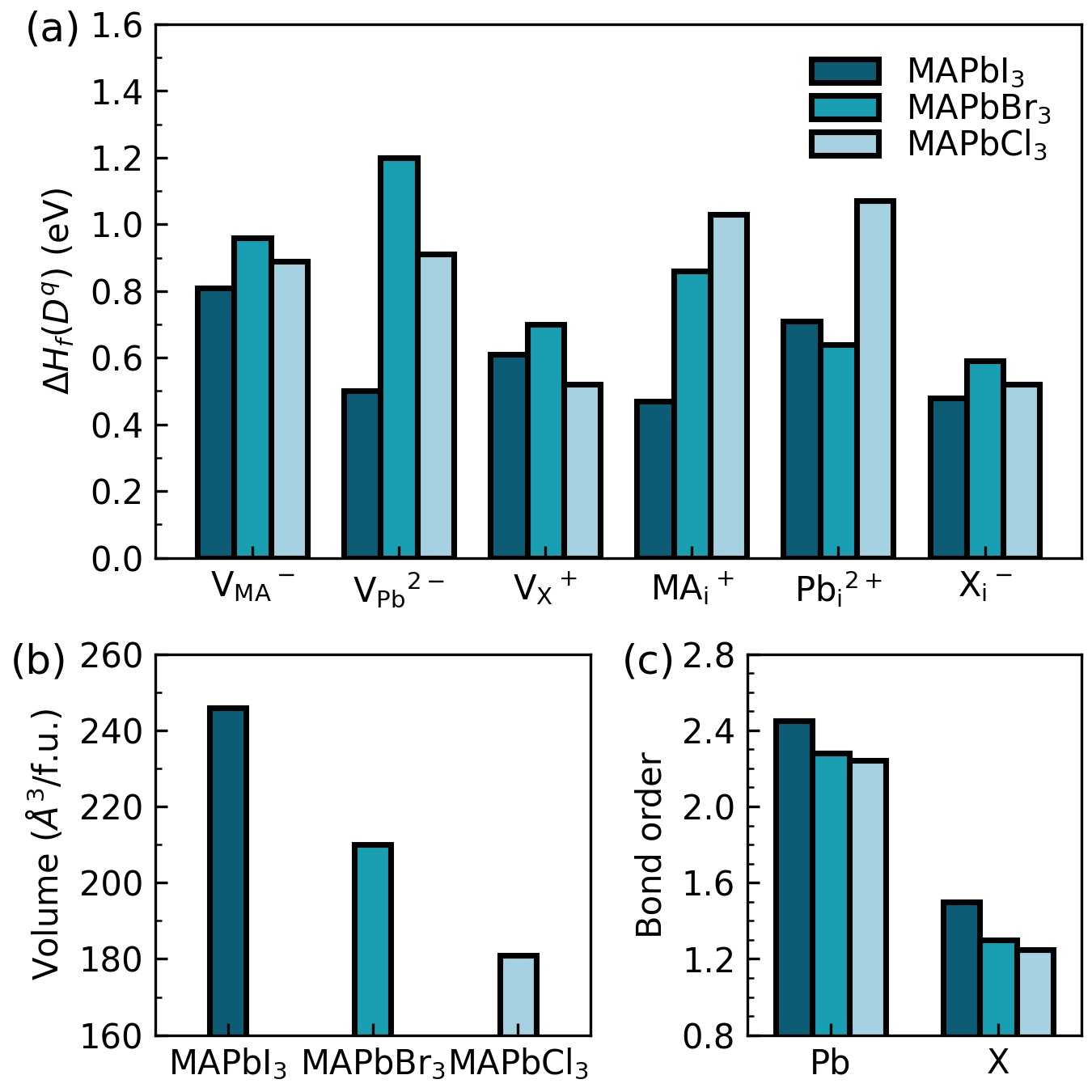}
    \caption{(a) Defect formation energies at intrinsic conditions of MAPbI$_3$, MAPbBr$_3$ and MAPbCl$_3$; (b) volume per formula unit of the pristine perovskite lattices; (c) bond orders of Pb and halide ions, where larger values indicate stronger covalent bonds \cite{Manz2016, Manz2017, Manz2016program}.} \label{fig: varying X anion}
\end{figure}

The DFEs of defects at intrinsic conditions in MAPbI$_3$, MAPbBr$_3$, and MAPbCl$_3$ are compared in Fig. \ref{fig: varying X anion}(a). A uniform trend is not present, but generally the DFEs increase going from I to Br and then decrease again going from Br to Cl. Exceptions to this trend are the MA interstitial, whose DFE increases going from I to Br to Cl, and the Pb interstitial, whose DFE decreases going from I to Br, before increasing again going from Br to Cl. 

The trend for the vacancies can be explained by two competing factors. The cell volume decreases significantly going from I to Br to Cl, as shown in Fig. \ref{fig: varying X anion}(b). This increases the electrostatic (Madelung) potential, and makes it energetically more difficult to create a vacancy in the lattice. However, also the chemical bond order of Pb cation and X anion decrease going from I to Br to Cl, as shown in Fig. \ref{fig: varying X anion}(c), which should make it easier to create a vacancy. The competition between these two factors gives a non-monotonic DFE increasing from I to Br, but decreasing from Br to Cl.

The volume effect works opposite for interstitials as compared to vacancies, which is mainly connected to the size of the interstitial. As the cell volume decreases going from I to Br to Cl, it becomes energetically more difficult to insert a large interstitial in the lattice. This is particularly true for the large MA interstitial, Fig. \ref{fig: varying X anion}(a), whose DFE increases going from I to Br to Cl. It is true to a somewhat lesser extend for the Pb interstitial, where the DFE in the I lattice is smaller than in the Cl lattice. The anomaly is the Br lattice, where the DFE of the Pb interstitial is relatively small. For the X interstitial, the DFE changes very little going from I to Br to Cl, Fig. \ref{fig: varying X anion}(a). Indeed, for this case there should be only a small volume effect, because, as the volume becomes smaller going from I to Br to Cl, so does the interstitial ion.

Among these defects, the variations in DFEs of vacancies $\mathrm{V_{MA}}^-$, $\mathrm{V_X}^+$ and the interstitial $\mathrm{X_i}^-$ are relatively moderate, $\lesssim$0.15 eV. These defects are less sensitive to the volume changes of the perovskite lattice, since the changes they induce in the lattice can be compensated to a certain degree by changes in I-Pb-I angles and the tilting of octahedra, as shown in the Supplemental Material Figs. S4 (c), (d) and (f). In contrast, the MA and Pb interstitials induce significant changes in the local structure, while the vacancy Pb involves significant breaking of Pb-X bonds, see Figs. S4 (a), (b) and (e). As a result, their DFEs have a larger variation of up to 0.7 eV upon changing the halogen X in the perovskites.

Due to its consistently small DFE, the halide interstitial $\mathrm{X}_i^-$ is always the most stable negatively charged defect in MAPbX$_3$ under intrinsic conditions. The situation is more varied for the compensating positively charged defects. Whereas for MAPbI$_3$ the dominant positively charged defect is the MA interstitial $\mathrm{MA}_i^+$, for MAPbBr$_3$ the dominant positively charged defect is the Pb interstitial $\mathrm{Pb_i}^{2+}$, and for MAPbCl$_3$ it is the halide vacancy $\mathrm{V_Cl}^+$, Fig. \ref{fig: DFEs and CSTLs}(a-c). 

To be specific, the two most stable defects at intrinsic conditions in MAPbI$_3$ are the interstitials $\mathrm{MA_i}^+$ and $\mathrm{I_i}^-$ with DFEs of 0.47 eV and 0.48 eV, respectively, Fig. \ref{fig: DFEs and CSTLs}(a), and equilibrium concentrations of 1.37$\times 10^{14}$ cm$^{-3}$ and 1.09$\times 10^{14}$ cm$^{-3}$ at room temperature, respectively, Eq. (\ref{eq:defect_concentration}). A third, but somewhat less prominent, defect is the vacancy $\mathrm{V_{Pb}}^{2-}$ with a DFE of 0.50 eV, and an equilibrium concentration of 1.44$\times 10^{13}$ cm$^{-3}$. All other defects have a DFE that is more than 0.1 eV larger than that of the two interstitials, leading to an equilibrium concentration that is more than two orders of magnitude smaller.

The dominant defects in MAPbBr$_3$ are $\mathrm{Pb_i}^{2+}$ and $\mathrm{Br_i}^-$ with DFEs of 0.64 eV and 0.59 eV, respectively, leading to equilibrium concentrations at room temperature of 1.03$\times 10^{12}$ cm$^{-3}$ and 2.07$\times 10^{12}$ cm$^{-3}$, respectively. Note that these concentrations are two orders of magnitude smaller than those in MAPbI$_3$, indicating that MAPbBr$_3$ is a more stable material. In MAPbCl$_3$, $\mathrm{V_{Cl}}^+$ and $\mathrm{Cl_i}^-$ are the most prominent defects, with a DFE and a concentration of 0.52 eV and 3.36$\times 10^{13}$ cm$^{-3}$, respectively. With a concentration roughly an order of magnitude larger than that in MAPbBr$_3$, MAPbCl$_3$ is somewhat less stable, though still more stable than MAPbI$_3$. It is interesting to note that only for MAPbCl$_3$ the dominant defects support the classic picture of Frenkel defects in ionic crystals, as represented by the halide vacancy and interstitial, which is consistent with the fact that the bonding in this material of the three is the most dominantly ionic. 

The CSTLs of MAPbI$_3$, MAPbBr$_3$, and MAPbCl$_3$ can be compared in Figs. \ref{fig: DFEs and CSTLs}(g)-(i). Although the SCAN+rVV10 functional is not designed to produce an accurate band gap, it is still possible to make some general statements about the electronic activity of the various defects. Under intrinsic conditions all defects in AMX$_3$ are charged, and positive/negative defects can act as potential traps for electrons/holes. 

Starting with MA-related defects, neither the top valence bands, nor the bottom conduction bands have any MA character \cite{Tao2019}, and the main effect of $\mathrm{MA_i}^+$ or $\mathrm{V_{MA}}^-$ defects is to create a local electrostatic potential that is attractive for electrons or holes, respectively. As electrostatic screening decreases going from the I to the Br to the Cl compound, one might expect this potential to become stronger and the associated defect states to become deeper. 

For instance, the CSTL $\mathrm{MA_i}$ (+/0) is only 0.05 eV below the CBM in MAPbI$_3$, making it a shallow electron trap. In MAPbBr$_3$ and MAPbCl$_3$ the CSTL $\mathrm{MA_i}$ (+/0) is 0.12 eV and 0.21 eV below the CBM, respectively. Whereas the latter are not shallow traps anymore, the DFEs of $\mathrm{MA_i}^+$ in MAPbBr$_3$ and MAPbCl$_3$ under intrinsic conditions are 0.86 eV and 1.03 eV, leading to concentrations ($4.28\times 10^7$ and $7.11\times 10^4$ cm$^{-3}$) that are negligible. In contrast, in MAPbI$_3$ the interstitial $\mathrm{MA_i}^+$ is one of the dominant effects, but there it only forms a shallow trap.

Proceeding with other relatively abundant defects in MAPbI$_3$, I interstitials and Pb vacancies, their CSTLs $\mathrm{I_i}$ ($-$/0) and $\mathrm{V_{Pb}}$ (2$-$/$-$) are close to the VBM. The fact that these defects do not form states that are deep inside the band gap, has a simple explanation. This is because the upper valence band has dominant I character, and consists of I-Pb anti-bonding states, with the maximally anti-bonding state at the VBM \cite{Tao2019}. Defect states associated with I interstitials and Pb vacancies also have I character. As bonding or anti-bonding at such defects is not larger than at the I ions in the pristine material, the associated defect states are not likely to have an energy above the VBM, even if negatively charging these localized states should push their energy upwards. This reasoning should hold for all AMX$_3$ compounds, and indeed it can be observed in Fig. \ref{fig: DFEs and CSTLs} that $\mathrm{X_i}$ ($-$/0) and $\mathrm{V_{Pb}}$ (2$-$/$-$) CSTL remain close to the VBM.

Finally, we consider the defects with electronic states that have Pb character, i.e. X vacancies and Pb interstitials. The lowest conduction band in APbX$_3$ has dominant Pb character, and consists of I-Pb anti-bonding states \cite{Tao2019}. Following the above reasoning, defects with Pb character involve non-optimal (anti-)bonding. This means that their states are likely to appear below the conduction band, or, in other words, inside the band gap of the semiconductor. Positively charging these localized states should push their energy even further downwards.

Indeed, defects in MAPbI$_3$ that give less shallow levels are $\mathrm{V_I}^+$ and $\mathrm{Pb_i}^{2+}$, which can act electron traps with (+/0) and (2+/+) transitions of 0.15 eV and 0.31 eV below the CBM, respectively. However, as the DFEs of these two defects under equilibrium conditions are 0.61 eV and 0.71 eV, respectively, their equilibrium concentrations of $6.10\times 10^{11}$ and $6.36\times 10^{10}$ cm$^{-3}$ are too small to make these defects harmful. 

\begin{figure}
    \includegraphics[width=0.7\columnwidth]{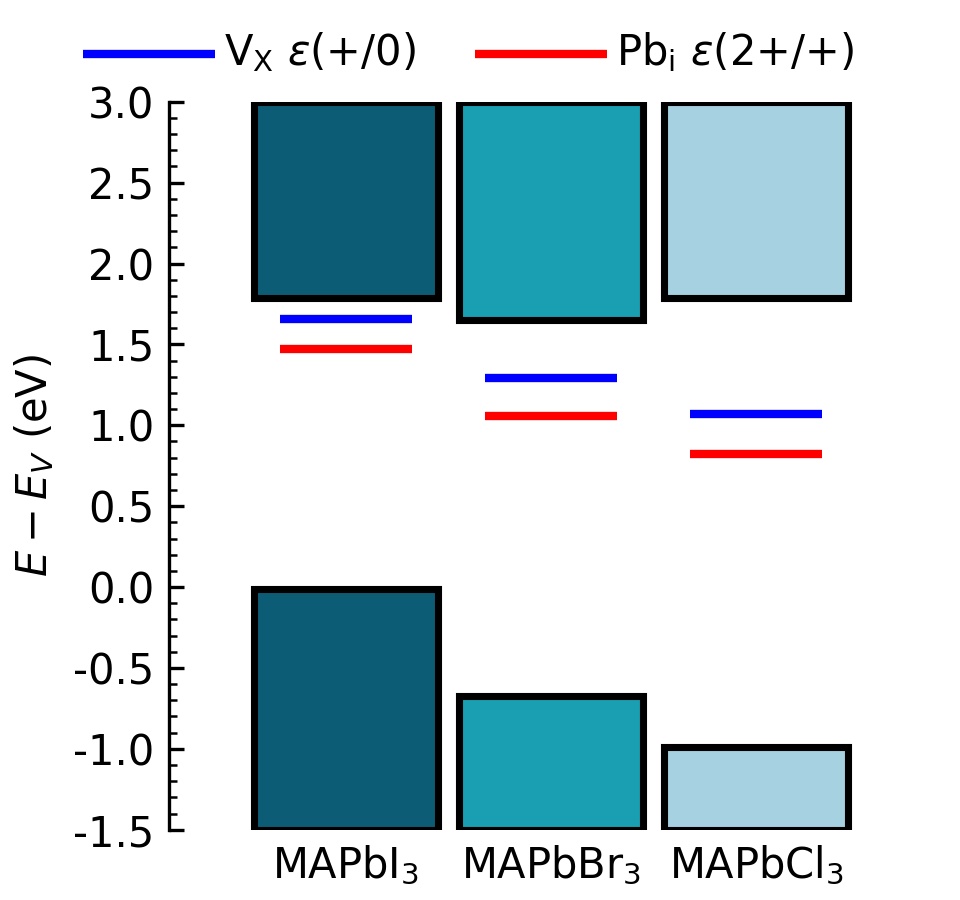}
    \caption{Charge state transition levels associated with $\mathrm{V_X}$ [$\varepsilon(+/0)$] and $\mathrm{Pb_i}$ [$\varepsilon(2+/+)$] defects in MAPbX$_3$, X = I, Br, Cl. The colored boxes indicate the relative positions of the valence and conduction bands, according to Ref. \onlinecite{Tao2019}. \label{fig: Deep transition levels}}
\end{figure}

The same is not true for the $\mathrm{Pb_i}^{2+}$ and $\mathrm{V_X}^+$ defects in MAPbBr$_3$ and MAPbCl$_3$. As shown in Fig. \ref{fig: Deep transition levels}, both these defects become deeper electron traps going from the I- to the Br- and the Cl-compound. This is potentially harmful, because $\mathrm{Pb_i}^{2+}$ is a dominant defect in MAPbBr$_3$, and $\mathrm{V_{Cl}}^+$ is a dominant defect in MAPbCl$_3$, as discussed above. 

In summary, considering both DFEs and CSTLs, there are no detrimental defects in MAPbI$_3$, as relatively abundant defects ($\mathrm{MA_i}^+$, $\mathrm{I_i}^-$ and $\mathrm{V_{Pb}}^{2-}$) only create shallow traps, and defects with less shallow levels ($\mathrm{V_I}^+$ and $\mathrm{Pb_i}^{2+}$) have a negligible concentration under equilibrium conditions, which agrees with the general consensus \cite{Yin2014, Meggiolaro2018review}. In contrast, $\mathrm{Pb_i}^{2+}$ and $\mathrm{V_{Cl}}^+$ are the most abundant defects in MAPbBr$_3$ and MAPbCl$_3$, respectively, and have deep levels that can act as electron traps. Our results therefore indicate that the latter materials are less defect tolerant than previous calculations seem to have indicated \cite{Buin2015,Shi2015a}. 
     
\subsection{Varying the A cation}\label{subsection: varying the A cation}

\begin{figure}
    \includegraphics[width=1\columnwidth]{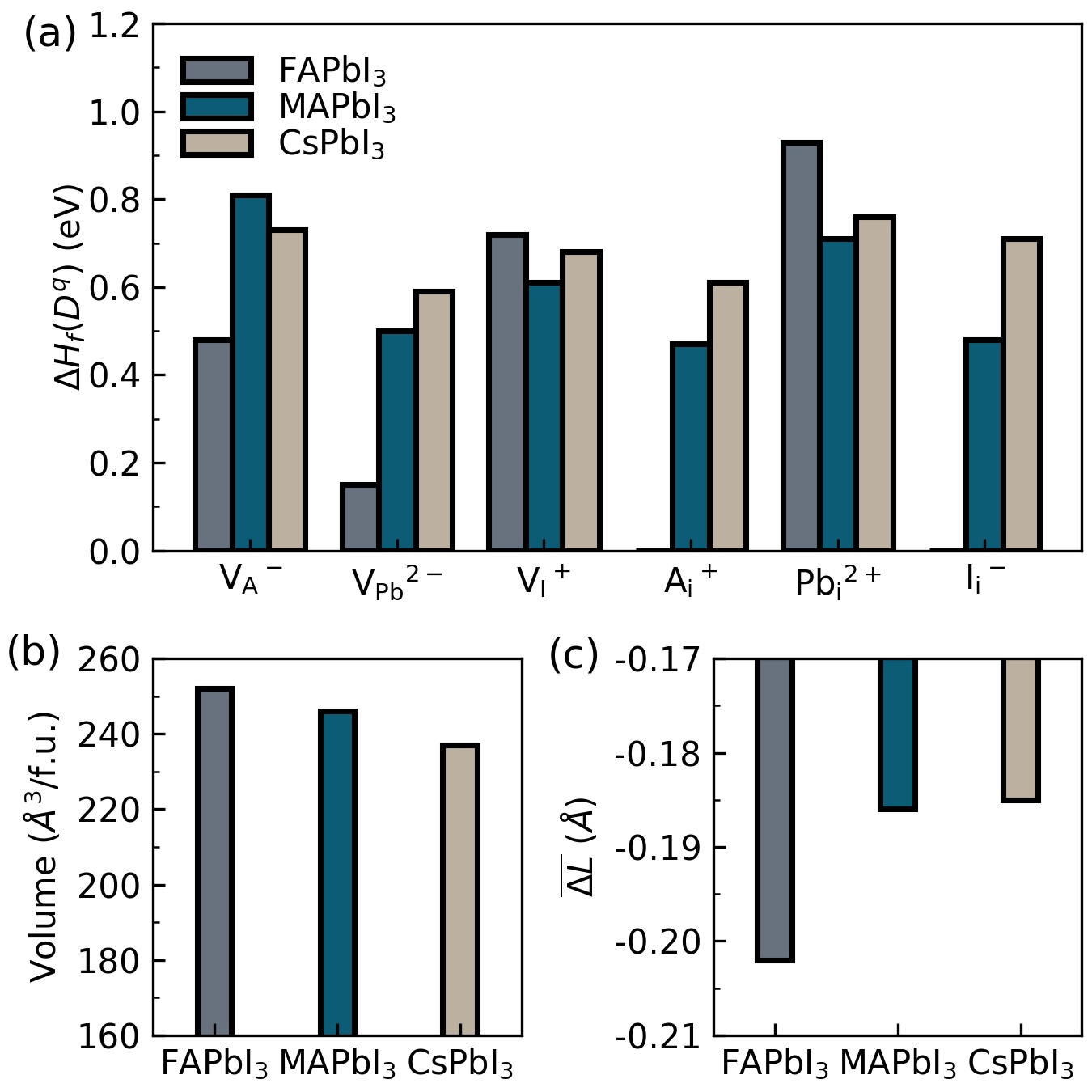}
    \caption{(a) Defect formation energies at intrinsic conditions of FAPbI$_3$, MAPbI$_3$ and CsPbI$_3$; (b) volume per formula unit of the pristine perovskite lattices; (c) average Pb-I bond length changes of the six I atoms closest to a Pb vacancy.}\label{fig: varying A cation}
\end{figure}

The DFEs at intrinsic conditions of defects in FAPbI$_3$, MAPbI$_3$, and CsPbI$_3$ are compared in Fig. \ref{fig: varying A cation}(a). Again, a uniform trend upon changing the A cation in the perovskite lattice is not present, but for four of the six defects, the DFEs increase significantly going from FA to MA, whereas going from MA to Cs the changes are much smaller. Exceptions to this trend are the I vacancy and the Pb interstitial, where the DFEs in the FA material are larger than those in the other two.

The general increase in DFEs going from FA to MA to Cs can be explained as a volume effect. Along this series the cell volume decreases, see Fig. \ref{fig: varying A cation}(b). This makes it more difficult to create a vacancy, as the Madelung potential becomes stronger. It also makes it more difficult to introduce an interstitial, as the lattice becomes more compact.

Noticeable is the rather extreme behavior of FAPbI$_3$, where the formation energies of $\mathrm{FA_i}^+$, and $\mathrm{I_i}^-$ are much smaller than their counterparts in MAPbI$_3$ and CsPbI$_3$, with DFEs of $\sim$0 eV. The ease with which these two interstitials can be introduced reflects the space in the lattice created by its rather large volume, Fig. \ref{fig: varying A cation}(b). 
In addition, this also attributes to that introducing FA and iodine interstitials can further release the strain in the FAPbI$_3$ lattice through promoting the octahedral tilting, as shown in Supplemental Material Fig. S1. Whereas, incorporating Pb interstitial is different, which will be explained below.

The vacancy $\mathrm{V_{Pb}}^{2-}$ is also fairly easy to create in FAPbI$_3$ with a DFE of 0.15 eV. We suggest that this is because this Pb vacancy allows for a local relaxation of the lattice of PbI6 octahedra, which is fairly strained in FAPbI$_3$, see Supplemental Material Fig. S1. A way to visualize this is to look at the changes in Pb-I bond lengths for the I atoms neighboring the Pb vacancy. As shown in Fig. \ref{fig: varying A cation}(c), compared to MAPbI$_3$ and CsPbI$_3$, a larger shortening of Pb-I bonds occurs in FAPbI$_3$, indicating a stronger structural relaxation and local octahedral tilting.

In contrast, the DFE of the interstitial $\mathrm{Pb_i}^{2+}$ in FAPbI$_3$ is larger than in MAPbI$_3$ and CsPbI$_3$. The Pb interstitial is embedded in the lattice through a fivefold coordination with I atoms of neighboring octahedra, see Supplemental Material Fig. S2(b), which is more difficult to achieve in the already more distorted and strained lattice of FAPbI$_3$. The other exception to the general trend mentioned above is the $\mathrm{V_I}^+$ vacancy, which has a similar DFE for all three compounds. Apparently the local bonding of an I atom in the lattice does not change very much in these compounds.

The dominant defects in FAPbI$_3$ are the same as the ones identified for MAPbI$_3$ in Sec. \ref{subsection: varying the X anion}, i.e., the interstitials $\mathrm{A_i}^+$ ($\mathrm{A} = \mathrm{FA,MA}$), and $\mathrm{I_i}^-$, and to a lesser extend the vacancy $\mathrm{V_{Pb}}^{2-}$. A major difference, however, is that the DFEs of these defects are much smaller for FAPbI$_3$ than for MAPbI$_3$. In fact, the DFEs of  $\mathrm{FA_i}^+$ and $\mathrm{I_i}^-$ are $\sim$0 eV, indicating that the material FAPbI$_3$ is intrinsically not very stable.

In CsPbI$_3$, the relative importance of the three types of defects mentioned above changes somewhat. Whereas the interstitial $\mathrm{A_i}^+$ ($\mathrm{A} = \mathrm{Cs}$) is still the dominant positively charged defect, the roles of the two negatively charged defects $\mathrm{I_i}^-$ and $\mathrm{V_{Pb}}^{2-}$ are interchanged, in the sense that the latter now has become dominant. Specifically, the DFEs of $\mathrm{Cs_i}^+$ and $\mathrm{V_{Pb}}^{2-}$ are 0.61 eV and 0.59 eV, respectively, giving rise to equilibrium concentrations of 8.08$\times 10^{11}$ cm$^{-3}$ and 4.32$\times 10^{11}$ cm$^{-3}$, respectively. The defect concentration in CsPbI$_3$ is two orders of magnitude lower than that in MAPbI$_3$, indicating better stability of the material with respect to defect formation. 

Comparing the CSTLs of MAPbI$_3$, FAPbI$_3$, and CsPbI$_3$, Figs. \ref{fig: DFEs and CSTLs} (g), (j), and (k), one observes that there is actually little difference between the materials. The analysis of the position and character of the CSTLs as presented in Sec. \ref{subsection: varying the X anion} still holds. This makes sense, as the A cation does not directly participate to the electronic structure of the valence and conduction bands around the band gap \cite{Tao2019}.

The A cation related defects, $\mathrm{A_i}^+$ and $\mathrm{V_A}^-$, A = MA, FA, Cs, have levels close to the CBM and the VBM, respectively, and the defects with electronic I character, $\mathrm{I_i}^-$ and $\mathrm{V_{Pb}}^{2-}$, have levels close to the VBM, all of which represent shallow traps. The defects with electronic Pb character, $\mathrm{Pb_i}^{2+}$ and $\mathrm{V_{I}}^{+}$ can create deep traps, but their concentrations under equilibrium conditions are low.

\subsection{Varying the M cation}\label{subsection: varying the M cation}

Upon substituting the M cation in in AMX$_3$, from MAPbI$_3$ to MASnI$_3$, the DFEs of all defects change drastically, as can be observed from comparing Figs. \ref{fig: DFEs and CSTLs}(a) and (f). It is evident from the latter figure that the negatively charged defects $\mathrm{V_{MA}}^-$ and $\mathrm{I_i}^-$, and in particular the vacancy $\mathrm{V_{Sn}}^{2-}$, have a very small formation energy. The high stability of these defects has also been observed in previous theoretical studies of Sn-based perovskites \cite{Liang2021, Meggiolaro2020a, Shi2017, Xu2014}. 

Formation of negatively charged defects in MASnI$_3$ is encouraged by the relatively low ionization potential of this compound \cite{Tao2019,Xiao2019,Meggiolaro2020a}. Holes to compensate for the negative charge of the defects are then easily formed in the valence band. The high energy of the valence band is ultimately due to the Sn 5s level being relatively high in energy, which correlates with the relative ease with which Sn(II) can be oxidized to Sn(IV) \cite{Tao2019}. In contrast, MAPbI$_3$ has a larger ionization potential, resulting from the Pb 6s level being relatively low in energy, correlating with a more difficult oxidation of Pb(II) to Pb(IV).

\begin{figure}
    \includegraphics[width=1\columnwidth]{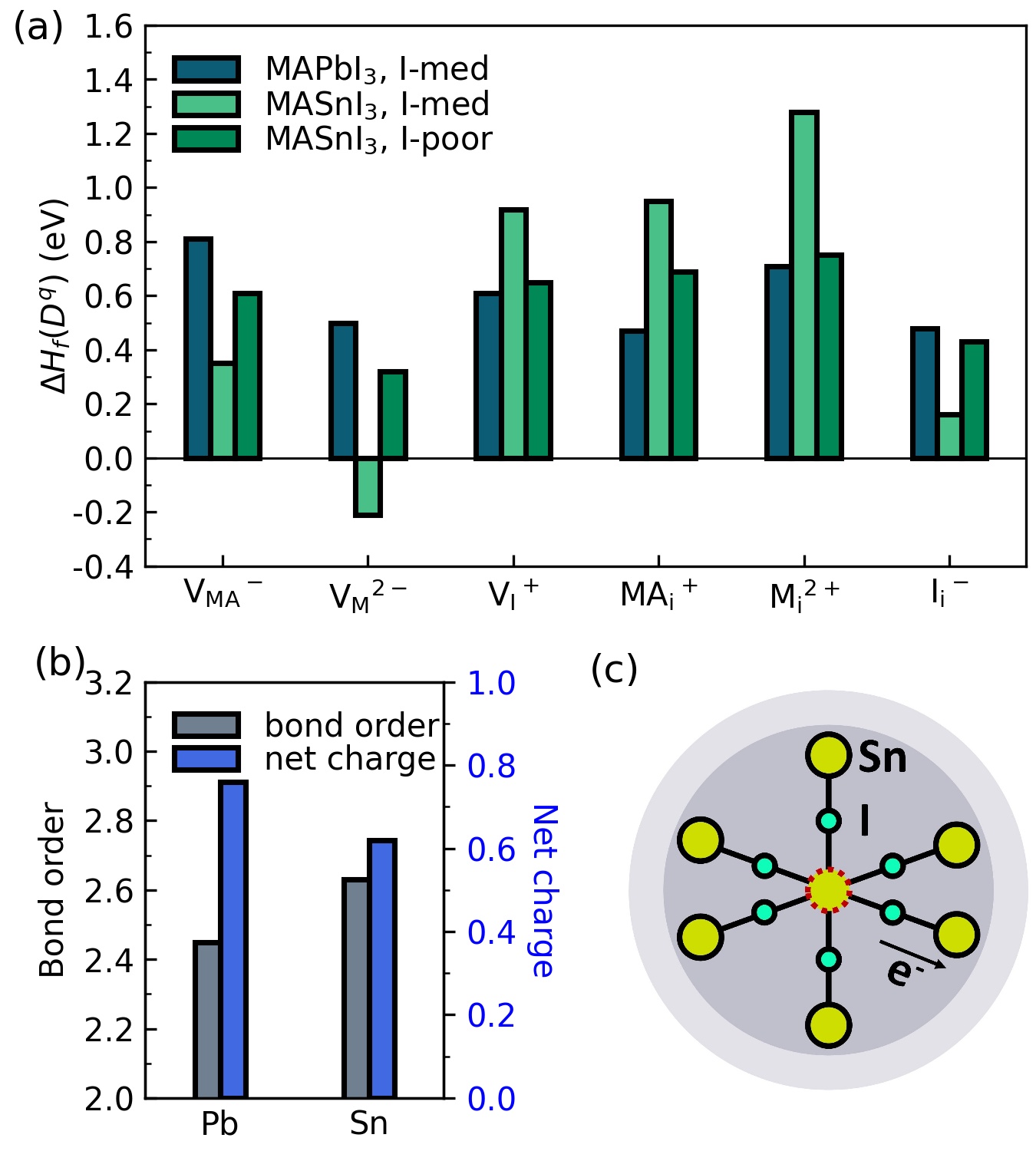}
    \caption{(a) Defect formation energies (DFEs) of MAPbI$_3$ and MASnI$_3$ under I-medium and I-poor conditions, see Fig. \ref{fig:chemicalpotentials} (the DFEs of MAPbI$_3$ are identical under both conditions, whereas those of MASnI$_3$ vary, see main text); (b) bond order of the M-I bond and net atomic charge on M in MAPbI$_3$ and MASnI$_3$; (c) diagram illustrating the charge spill-over upon creating a $\mathrm{V_{M}}^{2-}$ defect. }\label{fig: varying M cation}
\end{figure}

At the same time, the DFEs of positively charged defects in MASnI$_3$ remains large, so that the charge of these defects does not compensate that of the negatively charged defects. Charge neutrality then has to be achieved by the hole charge carriers in the perovskite, see Eq. \ref{eq:charge_neutrality}, which results in MASnI$_3$ being an intrinsic p-type semiconductor. The DFEs of MASnI$_3$ are shown in Fig. \ref{fig: varying M cation}, calculated under the assumption that the Fermi level is pinned at the top of the valence band, i.e., $E_F^{(i)}=0$. In fact, the vacancy $\mathrm{V_{Sn}}^{2-}$ still has a negative formation energy of $-0.21$ eV under these conditions, indicating that MASnI$_3$ is a degenerate p-type semiconductor. 

In order to avoid such a large amount of Sn vacancies, the growth conditions of the MASnI$_3$ perovskite can be tuned to iodine-poor conditions, for instance by increasing the relative amount of SnI$_2$ in the precursors \cite{Wang2014}. At iodine-poor conditions, the formation of $\mathrm{V_{Sn}}^{2-}$ defects can be significantly suppressed, with a DFE raised to 0.32 eV, Fig. \ref{fig: varying M cation}(a). However, also under these conditions MASnI$_3$ remains an intrinsic p-type semiconductor, see the Supplemental Material Fig. S5(c).  

The DFEs of MAPbI$_3$ are actually the same under I-medium and I-poor conditions, which stems from a compensation between the change in the iodine chemical potential, and the change in the intrinsic Fermi level, Sec. \ref{sec: DFE} \cite{Xue2021}. This compensation mechanism does not work if the Fermi level is not pinned, as is the case for MASnI$_3$. Comparing the DFEs of MAPbI$_3$ and MASnI$_3$ under I-poor conditions, which are most physical for MASnI$_3$, we observe that there is only a moderate difference. The DFEs of $\mathrm{V_{I}}^+$, $\mathrm{I_i}^-$, and $\mathrm{M_i}^{2+}$ differ by $\lesssim$0.05 eV between the two compounds.

In Secs. \ref{subsection: varying the X anion} and \ref{subsection: varying the A cation} changes in equilibrium volume of the different perovskites have been involved to explain the trends in DFEs. As the equilibrium volume of MASnI$_3$ is only 2\% smaller than that of MAPbI$_3$, see Supplemental Material Table S2, volume effects are small.

The DFEs of $\mathrm{V_{MA}}^-$ and $\mathrm{MA_i}^+$ in MASnI$_3$ are $\sim$0.2 eV lower, respectively higher, than in MAPbI$_3$. It can be explained by the fact that the interaction of the MA cations with the SnI framework in MASnI$_3$ is weaker than with the PbI framework in MAPbI$_3$, which facilitates the formation of MA vacancies in the former, and impedes the formation of MA interstitials. The Sn-I bond is slightly more covalent than the Pb-I, as demonstrated by the larger bond order, and the smaller charge on the Sn versus the Pb ions, Fig. \ref{fig: varying M cation}(b). The charges on the I ions in the SnI framework are then also smaller than in the PbI framework, thus their Coulomb interaction with the MA cations is somewhat weaker. 

The DFE of $\mathrm{V_{Sn}}^{2-}$ in MASnI$_3$ is $\sim$0.2 eV lower than $\mathrm{V_{Pb}}^{2-}$ in MAPbI$_3$. As discussed above, the ease with which Sn vacancies are formed is facilitated by the special chemistry of Sn. The $2-$ charge on $\mathrm{V_{Sn}}$ in MASnI$_3$ is more localized than that of $\mathrm{V_{Pb}}$ in MAPbI$_3$. The charge on the Sn atoms in a shell surrounding the Sn vacancy, see Fig. \ref{fig: varying M cation}(c), is in total reduced by 0.16$e$, whereas for Pb atoms the reduction is 0.27$e$. Sn more easily retains its positive charge, or put differently, it is more easily oxidized.

The positions of the CSTLs with respect to the band edges in MASnI$_3$ are similar to those in MAPbI$_3$. Defects with I character, such as I interstitials and Sn vacancies, give only shallow states just above the VBM, for which we suggest an explanation similar to that given in Sec. \ref{subsection: varying the X anion}. One concludes that, although Sn vacancies can be relatively abundant, they do not give rise to trap states for holes. Defects with Sn character, such as Sn interstitials and I vacancies, give states below the CBM, where in particular the $\mathrm{Sn_i}$ $(2+/+)$ level forms a deep trap for electrons. However, Sn interstitials are rare under equilibrium conditions, so their electronic effect is minimal. The MA related defects have no important electronic consequences, either because they only give shallow levels ($\mathrm{MA_i}$), or because their concentration is low ($\mathrm{V_{MA}}$).

In summary, the defect chemistry of MASnI$_3$ is rather special, resulting in a very high equilibrium concentration $1.45\times 10^{16}$ cm$^{-3}$ of $\mathrm{V_{Sn}}^{2-}$ defects, (at I-poor conditions). The charge in these vacancies is compensated mainly by holes in the valence band, making this material a degenerate intrinsic p-type semiconductor. In contrast to the significant change in defect concentrations, upon changing the M cation from Pb to Sn, the CSTLs show no qualitative changes. Defects that are abundant under equilibrium conditions only display shallow acceptor levels, and defects that can act as electron traps appear only at low concentrations.   

\section{Summary and conclusions} \label{section: conclusions}

\begin{table} 
    \caption{Dominant defects and major source of deep traps in different perovskites.}
    \label{table: dominant defects}
    \begin{ruledtabular}
        \begin{tabular}{p{0.2\columnwidth} p{0.35\columnwidth} p{0.3\columnwidth}}
        Perovskites & Dominant defects & Major source of deep traps\\
        \hline
    MAPbI$_3$ & $\mathrm{MA_i}^+$, $\mathrm{I_i}^-$, $\mathrm{V_{Pb}}^{2-}$ & None \\
    MAPbBr$_3$ & $\mathrm{Pb_i}^{2+}$, $\mathrm{V_{Br}}^+$, $\mathrm{Br_i}^-$ & $\mathrm{Pb_i}^{2+}$, $\mathrm{V_{Br}}^+$ \\
    MAPbCl$_3$ & $\mathrm{V_{Cl}}^+$, $\mathrm{Cl_i}^-$ & $\mathrm{V_{Cl}}^+$ \\
    FAPbI$_3$ & $\mathrm{FA_i}^+$, $\mathrm{I_i}^-$ & None \\
    CsPbI$_3$ & $\mathrm{Cs_i}^+$, $\mathrm{V_{Pb}}^{2-}$ & None \\
    MASnI$_3$ &  $\mathrm{V_{Sn}}^{2-}$ & None \\
    \end{tabular}
    \end{ruledtabular}
\end{table}

\begin{table*} 
    \caption{Formation energies $\Delta H_f$ (eV) and concentrations $c$ (cm$^{-3}$) at $T$ = 300 K  of charged defects in six perovskites calculated using the SCAN+rVV10 functional. The results of MASnI$_3$ are calculated based upon I-poor conditions, while those of the other perovskites are calculated based on halide-medium conditions. Formation energies and concentrations of the dominant defects in each perovskite are underlined. Concentrations of defects in FAPbI$_3$ are not given because the perovskite FAPbI$_3$ is intrinsically not very stable.}
    \label{table: DFE of charged defects}
    \begin{ruledtabular}
    \begin{tabular}{lllllll}
        Perovskites & $\mathrm{V_{A}}^-$ & $\mathrm{V_{M}}^{2-}$ & $\mathrm{V_{X}}^+$ & $\mathrm{A_i}^+$ & $\mathrm{M_i}^{2+}$ & $\mathrm{X_i}^-$\\
        \hline
    \multicolumn{7}{l}{Defect formation energy $\Delta H_f$ (eV)}\\
    MAPbI$_3$ & 0.81 & \underline{0.50} & 0.61 & \underline{0.47} & 0.71 & \underline{0.48} \\
    MAPbBr$_3$ & 0.96 & 1.20 & 0.70 & 0.86 & \underline{0.64} & \underline{0.59} \\
    MAPbCl$_3$ & 0.89 & 0.91 & \underline{0.52} & 1.03 & 1.07 & \underline{0.52} \\
    FAPbI$_3$ & 0.48 & 0.15 & 0.72 & \underline{0.00} & 0.93 & \underline{0.00} \\
    CsPbI$_3$ & 0.73 & \underline{0.59} & 0.68 & \underline{0.61} & 0.76 & 0.71 \\
    MASnI$_3$ & 0.61 & \underline{0.32} & 0.65 & 0.69 & 0.75 & 0.43 \\
    \multicolumn{7}{l}{Defect concentration $c$ (cm$^{-3}$)}\\
    MAPbI$_3$ & 8.68$\times 10^{7}$ & \underline{1.44$\times 10^{13}$} & 6.10$\times 10^{11}$ & \underline{1.37$\times 10^{14}$} & 6.36$\times 10^{10}$ & \underline{1.09$\times 10^{14}$} \\
    MAPbBr$_3$ & 3.46$\times 10^{5}$ & 3.91$\times 10^{1}$ & 2.28$\times 10^{10}$ & 4.28$\times 10^{7}$ & \underline{1.03$\times 10^{12}$} & \underline{2.07$\times 10^{12}$} \\
    MAPbCl$_3$ & 5.79$\times 10^{6}$ & 2.96$\times 10^{6}$ & \underline{3.36$\times 10^{13}$} & 7.11$\times 10^{4}$ & 7.76$\times 10^{4}$ & \underline{3.36$\times 10^{13}$} \\
    CsPbI$_3$ & 2.00$\times 10^{9}$ & \underline{4.32$\times 10^{11}$} & 5.32$\times 10^{10}$ & \underline{8.08$\times 10^{11}$} & 9.55$\times 10^{9}$ & 1.54$\times 10^{10}$ \\
    MASnI$_3$ & 1.99$\times 10^{11}$ & \underline{1.45$\times 10^{16}$} & 1.39$\times 10^{11}$ & 3.52$\times 10^{10}$ & 1.20$\times 10^{10}$ & 7.08$\times 10^{14}$ \\
    \end{tabular}
    \end{ruledtabular}
\end{table*}

We have conducted a systematic investigation of the defect formation energies (DFEs) and charge state transition levels (CSTLs) of the intrinsic point defects, vacancies and interstitials, in six primary metal halide perovskites, MAPbI$_3$, MAPbBr$_3$, MAPbCl$_3$, FAPbI$_3$, CsPbI$_3$ and MASnI$_3$. The accurate and efficient SCAN+rVV10 functional is employed in all DFT calculations, and all structures are optimized self-consistently. Using MAPbI$_3$ as reference material, we have analysed the impact of changing anions and cations on the DFEs and CSTLs and identify the underlying physical origins of the observed trends. 

The dominant defects and major sources of deep traps are summarized in Table \ref{table: dominant defects}. The formation energies and concentrations of defects are summarized in Table \ref{table: DFE of charged defects}, and those of the neutral defects are given in the Supplemental Material, Table S3.

The changes in the DFEs upon replacing the halide anion in MAPbX$_3$, from X = I to Br to Cl, can be explained from a competition between two factors. Firstly, the lattice volume decreases, which increases the electrostatic (Madelung) potential in the lattice. Such increased electrostatic potential makes it easier to create (charged) interstitials, and more difficult to create vacancies. Secondly, the Pb-X bond order decreases from I to Br to Cl, which has the opposite effect.   

By varying the A cation in APbI$_3$, from A = FA to MA to Cs, one observes a general increase in the DFEs, which can be explained by a similar volume effect as changing the halogens. A smaller volume increases the Madelung potential, making it more difficult to create vacancies. In this case, a smaller volume also makes it more difficult to create an interstitial, as the lattice becomes more compact. In contrast to the case discussed in the previous paragraph, the chemical bonding is not much affected by A cation substitution when comparing Cs and MA. However, defects are easily created in FAPbI$_3$, where its large lattice volume allows for an easy creation of interstitials, and a large relaxation in the presence of vacancies.

Comparing MAPbI$_3$ and MASnI$_3$, Sn vacancies are much easier to form than Pb vacancies. So much, that $\mathrm{V_{Sn}}^{2-}$ is the only dominant defect in MASnI$_3$, which can only be charge compensated by holes in the valence band, making this compound an intrinsic p-type semiconductor.  

In these substitutions of anions and cations, the majority of CSTLs remain shallow, and move along with band edges. A-cation vacancies and interstitials do not interfere with the basic electronic structure of the perovskites. They only modify the local Coulomb potential slightly and hence form only shallow traps for holes and electrons. Defects that might catch holes, $\mathrm{X_i}^{-}$ and $\mathrm{V_{Pb}}^{2-}$, only create shallow levels around the VBM. The VBM has X character originating from maximally anti-bonding M-X states \cite{Tao2019}. The two mentioned defects also create states with X character, but with less anti-bonding character, and thus with a relatively low energy, even if negative charging pushes the defect levels upwards. 

In contrast, defects that might catch electrons, $\mathrm{M_i}^{2+}$ and $\mathrm{V_{X}}^{+}$, can create deep traps. The CBM has M-character originating from anti-bonding M-X states\cite{Tao2019}. The two defects create states with M character, but with less anti-bonding character, and thus with an energy below the CBM, which can become even lower upon the positive charging. The CSTLs generated by Pb interstitials and halide vacancies upon varying the X anion from I to Br and to Cl, can be classified as deep traps. Although these trap states are harmful electronically, the equilibrium concentrations of these defects in MAPbI$_3$ are negligible. However, in MAPbBr$_3$ and MAPbCl$_3$ the concentrations of these electron traps are higher, so that they are potentially harmful to the electronic properties.

\begin{acknowledgments}
H. Xue acknowledges the funding from the China Scholarship Council (CSC)(No. 201806420038). S. Tao acknowledges funding by the Computational Sciences for Energy Research (CSER) tenure track program of Shell and NWO (Project number 15CST04-2) and the NWO START-UP grant from the Netherlands.
\end{acknowledgments}

\bibliography{Manuscript}

\begin{thebibliography}{70}
\expandafter\ifx\csname natexlab\endcsname\relax\def\natexlab#1{#1}\fi
\expandafter\ifx\csname bibnamefont\endcsname\relax
  \def\bibnamefont#1{#1}\fi
\expandafter\ifx\csname bibfnamefont\endcsname\relax
  \def\bibfnamefont#1{#1}\fi
\expandafter\ifx\csname citenamefont\endcsname\relax
  \def\citenamefont#1{#1}\fi
\expandafter\ifx\csname url\endcsname\relax
  \def\url#1{\texttt{#1}}\fi
\expandafter\ifx\csname urlprefix\endcsname\relax\def\urlprefix{URL }\fi
\providecommand{\bibinfo}[2]{#2}
\providecommand{\eprint}[2][]{\url{#2}}

\bibitem[{\citenamefont{Jena et~al.}(2019)\citenamefont{Jena, Kulkarni, and
  Miyasaka}}]{Jena2019}
\bibinfo{author}{\bibfnamefont{A.~K.} \bibnamefont{Jena}},
  \bibinfo{author}{\bibfnamefont{A.}~\bibnamefont{Kulkarni}}, \bibnamefont{and}
  \bibinfo{author}{\bibfnamefont{T.}~\bibnamefont{Miyasaka}},
  \bibinfo{journal}{Chem. Rev.} \textbf{\bibinfo{volume}{119}},
  \bibinfo{pages}{3036} (\bibinfo{year}{2019}),
  \urlprefix\url{https://doi.org/10.1021/acs.chemrev.8b00539}.

\bibitem[{\citenamefont{Saparov and Mitzi}(2016)}]{Saparov2016}
\bibinfo{author}{\bibfnamefont{B.}~\bibnamefont{Saparov}} \bibnamefont{and}
  \bibinfo{author}{\bibfnamefont{D.~B.} \bibnamefont{Mitzi}},
  \bibinfo{journal}{Chem. Rev.} \textbf{\bibinfo{volume}{116}},
  \bibinfo{pages}{4558} (\bibinfo{year}{2016}), ISSN \bibinfo{issn}{15206890},
  \urlprefix\url{https://pubs.acs.org/doi/10.1021/acs.chemrev.5b00715}.

\bibitem[{\citenamefont{Stoumpos et~al.}(2013)\citenamefont{Stoumpos,
  Malliakas, and Kanatzidis}}]{Stoumpos2013}
\bibinfo{author}{\bibfnamefont{C.~C.} \bibnamefont{Stoumpos}},
  \bibinfo{author}{\bibfnamefont{C.~D.} \bibnamefont{Malliakas}},
  \bibnamefont{and} \bibinfo{author}{\bibfnamefont{M.~G.}
  \bibnamefont{Kanatzidis}}, \bibinfo{journal}{Inorg. Chem.}
  \textbf{\bibinfo{volume}{52}}, \bibinfo{pages}{9019} (\bibinfo{year}{2013}),
  ISSN \bibinfo{issn}{00201669},
  \urlprefix\url{https://pubs.acs.org/doi/10.1021/ic401215x}.

\bibitem[{\citenamefont{Kojima et~al.}(2009)\citenamefont{Kojima, Teshima,
  Shirai, and Miyasaka}}]{Kojima2009}
\bibinfo{author}{\bibfnamefont{A.}~\bibnamefont{Kojima}},
  \bibinfo{author}{\bibfnamefont{K.}~\bibnamefont{Teshima}},
  \bibinfo{author}{\bibfnamefont{Y.}~\bibnamefont{Shirai}}, \bibnamefont{and}
  \bibinfo{author}{\bibfnamefont{T.}~\bibnamefont{Miyasaka}},
  \bibinfo{journal}{J. Am. Chem. Soc.} \textbf{\bibinfo{volume}{131}},
  \bibinfo{pages}{6050} (\bibinfo{year}{2009}),
  \urlprefix\url{https://doi.org/10.1021/ja809598r}.

\bibitem[{\citenamefont{{National Renewable Energy
  Laboratory}}(2021)}]{NREL2021}
\bibinfo{author}{\bibnamefont{{National Renewable Energy Laboratory}}},
  \bibinfo{journal}{Best Research Cell Efficiencies}  (\bibinfo{year}{2021}),
  \urlprefix\url{https://www.nrel.gov/pv/cell-efficiency.html}.

\bibitem[{\citenamefont{Juarez-Perez et~al.}(2019)\citenamefont{Juarez-Perez,
  Ono, Uriarte, Cocinero, and Qi}}]{Juarez-Perez2019}
\bibinfo{author}{\bibfnamefont{E.~J.} \bibnamefont{Juarez-Perez}},
  \bibinfo{author}{\bibfnamefont{L.~K.} \bibnamefont{Ono}},
  \bibinfo{author}{\bibfnamefont{I.}~\bibnamefont{Uriarte}},
  \bibinfo{author}{\bibfnamefont{E.~J.} \bibnamefont{Cocinero}},
  \bibnamefont{and} \bibinfo{author}{\bibfnamefont{Y.}~\bibnamefont{Qi}},
  \bibinfo{journal}{ACS Appl. Mater. Interfaces} \textbf{\bibinfo{volume}{11}},
  \bibinfo{pages}{12586} (\bibinfo{year}{2019}), ISSN \bibinfo{issn}{19448252},
  \urlprefix\url{https://pubs.acs.org/doi/10.1021/acsami.9b02374}.

\bibitem[{\citenamefont{Xu et~al.}(2018)\citenamefont{Xu, Li, Jin, Liu, Bao,
  O'Carroll, and Tang}}]{Xu2018}
\bibinfo{author}{\bibfnamefont{R.~P.} \bibnamefont{Xu}},
  \bibinfo{author}{\bibfnamefont{Y.~Q.} \bibnamefont{Li}},
  \bibinfo{author}{\bibfnamefont{T.~Y.} \bibnamefont{Jin}},
  \bibinfo{author}{\bibfnamefont{Y.~Q.} \bibnamefont{Liu}},
  \bibinfo{author}{\bibfnamefont{Q.~Y.} \bibnamefont{Bao}},
  \bibinfo{author}{\bibfnamefont{C.}~\bibnamefont{O'Carroll}},
  \bibnamefont{and} \bibinfo{author}{\bibfnamefont{J.~X.} \bibnamefont{Tang}},
  \bibinfo{journal}{ACS Appl. Mater. Interfaces} \textbf{\bibinfo{volume}{10}},
  \bibinfo{pages}{6737} (\bibinfo{year}{2018}), ISSN \bibinfo{issn}{19448252},
  \urlprefix\url{https://pubs.acs.org/doi/10.1021/acsami.7b18389}.

\bibitem[{\citenamefont{Song et~al.}(2018)\citenamefont{Song, Wang, Phillips,
  Grice, Zhao, Yu, Chen, Li, Yin, Ellingson et~al.}}]{Song2018}
\bibinfo{author}{\bibfnamefont{Z.}~\bibnamefont{Song}},
  \bibinfo{author}{\bibfnamefont{C.}~\bibnamefont{Wang}},
  \bibinfo{author}{\bibfnamefont{A.~B.} \bibnamefont{Phillips}},
  \bibinfo{author}{\bibfnamefont{C.~R.} \bibnamefont{Grice}},
  \bibinfo{author}{\bibfnamefont{D.}~\bibnamefont{Zhao}},
  \bibinfo{author}{\bibfnamefont{Y.}~\bibnamefont{Yu}},
  \bibinfo{author}{\bibfnamefont{C.}~\bibnamefont{Chen}},
  \bibinfo{author}{\bibfnamefont{C.}~\bibnamefont{Li}},
  \bibinfo{author}{\bibfnamefont{X.}~\bibnamefont{Yin}},
  \bibinfo{author}{\bibfnamefont{R.~J.} \bibnamefont{Ellingson}},
  \bibnamefont{et~al.}, \bibinfo{journal}{Sustain. Energy Fuels}
  \textbf{\bibinfo{volume}{2}}, \bibinfo{pages}{2460} (\bibinfo{year}{2018}),
  ISSN \bibinfo{issn}{23984902},
  \urlprefix\url{http://xlink.rsc.org/?DOI=C8SE00358K}.

\bibitem[{\citenamefont{Juarez-Perez et~al.}(2018)\citenamefont{Juarez-Perez,
  Ono, Maeda, Jiang, Hawash, and Qi}}]{Juarez-Perez2018}
\bibinfo{author}{\bibfnamefont{E.~J.} \bibnamefont{Juarez-Perez}},
  \bibinfo{author}{\bibfnamefont{L.~K.} \bibnamefont{Ono}},
  \bibinfo{author}{\bibfnamefont{M.}~\bibnamefont{Maeda}},
  \bibinfo{author}{\bibfnamefont{Y.}~\bibnamefont{Jiang}},
  \bibinfo{author}{\bibfnamefont{Z.}~\bibnamefont{Hawash}}, \bibnamefont{and}
  \bibinfo{author}{\bibfnamefont{Y.}~\bibnamefont{Qi}}, \bibinfo{journal}{J.
  Mater. Chem. A} \textbf{\bibinfo{volume}{6}}, \bibinfo{pages}{9604}
  (\bibinfo{year}{2018}), ISSN \bibinfo{issn}{20507496},
  \urlprefix\url{http://xlink.rsc.org/?DOI=C8TA03501F}.

\bibitem[{\citenamefont{{Chun-Ren Ke} et~al.}(2017)\citenamefont{{Chun-Ren Ke},
  Walton, Lewis, Tedstone, O'Brien, Thomas, and Flavell}}]{Chun-RenKe2017}
\bibinfo{author}{\bibfnamefont{J.}~\bibnamefont{{Chun-Ren Ke}}},
  \bibinfo{author}{\bibfnamefont{A.~S.} \bibnamefont{Walton}},
  \bibinfo{author}{\bibfnamefont{D.~J.} \bibnamefont{Lewis}},
  \bibinfo{author}{\bibfnamefont{A.}~\bibnamefont{Tedstone}},
  \bibinfo{author}{\bibfnamefont{P.}~\bibnamefont{O'Brien}},
  \bibinfo{author}{\bibfnamefont{A.~G.} \bibnamefont{Thomas}},
  \bibnamefont{and} \bibinfo{author}{\bibfnamefont{W.~R.}
  \bibnamefont{Flavell}}, \bibinfo{journal}{Chem. Commun.}
  \textbf{\bibinfo{volume}{53}}, \bibinfo{pages}{5231} (\bibinfo{year}{2017}),
  ISSN \bibinfo{issn}{1364548X},
  \urlprefix\url{http://xlink.rsc.org/?DOI=C7CC01538K}.

\bibitem[{\citenamefont{Latini et~al.}(2017)\citenamefont{Latini, Gigli, and
  Ciccioli}}]{Latini2017}
\bibinfo{author}{\bibfnamefont{A.}~\bibnamefont{Latini}},
  \bibinfo{author}{\bibfnamefont{G.}~\bibnamefont{Gigli}}, \bibnamefont{and}
  \bibinfo{author}{\bibfnamefont{A.}~\bibnamefont{Ciccioli}},
  \bibinfo{journal}{Sustain. Energy Fuels} \textbf{\bibinfo{volume}{1}},
  \bibinfo{pages}{1351} (\bibinfo{year}{2017}), ISSN \bibinfo{issn}{23984902},
  \urlprefix\url{http://xlink.rsc.org/?DOI=C7SE00114B}.

\bibitem[{\citenamefont{Juarez-Perez et~al.}(2016)\citenamefont{Juarez-Perez,
  Hawash, Raga, Ono, and Qi}}]{Juarez-Perez2016}
\bibinfo{author}{\bibfnamefont{E.~J.} \bibnamefont{Juarez-Perez}},
  \bibinfo{author}{\bibfnamefont{Z.}~\bibnamefont{Hawash}},
  \bibinfo{author}{\bibfnamefont{S.~R.} \bibnamefont{Raga}},
  \bibinfo{author}{\bibfnamefont{L.~K.} \bibnamefont{Ono}}, \bibnamefont{and}
  \bibinfo{author}{\bibfnamefont{Y.}~\bibnamefont{Qi}},
  \bibinfo{journal}{Energy Environ. Sci.} \textbf{\bibinfo{volume}{9}},
  \bibinfo{pages}{3406} (\bibinfo{year}{2016}), ISSN \bibinfo{issn}{17545706},
  \urlprefix\url{http://xlink.rsc.org/?DOI=C6EE02016J}.

\bibitem[{\citenamefont{Jung et~al.}(2016)\citenamefont{Jung, Lee, Lee, Park,
  Raga, Ono, Wang, Leyden, Yu, Hong et~al.}}]{Jung2016}
\bibinfo{author}{\bibfnamefont{M.~C.} \bibnamefont{Jung}},
  \bibinfo{author}{\bibfnamefont{Y.~M.} \bibnamefont{Lee}},
  \bibinfo{author}{\bibfnamefont{H.~K.} \bibnamefont{Lee}},
  \bibinfo{author}{\bibfnamefont{J.}~\bibnamefont{Park}},
  \bibinfo{author}{\bibfnamefont{S.~R.} \bibnamefont{Raga}},
  \bibinfo{author}{\bibfnamefont{L.~K.} \bibnamefont{Ono}},
  \bibinfo{author}{\bibfnamefont{S.}~\bibnamefont{Wang}},
  \bibinfo{author}{\bibfnamefont{M.~R.} \bibnamefont{Leyden}},
  \bibinfo{author}{\bibfnamefont{B.~D.} \bibnamefont{Yu}},
  \bibinfo{author}{\bibfnamefont{S.}~\bibnamefont{Hong}}, \bibnamefont{et~al.},
  \bibinfo{journal}{Appl. Phys. Lett.} \textbf{\bibinfo{volume}{108}},
  \bibinfo{pages}{73901} (\bibinfo{year}{2016}),
  \urlprefix\url{http://aip.scitation.org/doi/10.1063/1.4941994}.

\bibitem[{\citenamefont{Dualeh et~al.}(2014)\citenamefont{Dualeh, Gao, Seok,
  Nazeeruddin, and Gr{\"{a}}tzel}}]{Dualeh2014}
\bibinfo{author}{\bibfnamefont{A.}~\bibnamefont{Dualeh}},
  \bibinfo{author}{\bibfnamefont{P.}~\bibnamefont{Gao}},
  \bibinfo{author}{\bibfnamefont{S.~I.} \bibnamefont{Seok}},
  \bibinfo{author}{\bibfnamefont{M.~K.} \bibnamefont{Nazeeruddin}},
  \bibnamefont{and}
  \bibinfo{author}{\bibfnamefont{M.}~\bibnamefont{Gr{\"{a}}tzel}},
  \bibinfo{journal}{Chem. Mater.} \textbf{\bibinfo{volume}{26}},
  \bibinfo{pages}{6160} (\bibinfo{year}{2014}), ISSN \bibinfo{issn}{15205002},
  \urlprefix\url{https://pubs.acs.org/doi/10.1021/cm502468k}.

\bibitem[{\citenamefont{Doherty et~al.}(2020)\citenamefont{Doherty, Winchester,
  Macpherson, Johnstone, Pareek, Tennyson, Kosar, Kosasih, Anaya, Abdi-Jalebi
  et~al.}}]{Doherty2020}
\bibinfo{author}{\bibfnamefont{T.~A.~S.} \bibnamefont{Doherty}},
  \bibinfo{author}{\bibfnamefont{A.~J.} \bibnamefont{Winchester}},
  \bibinfo{author}{\bibfnamefont{S.}~\bibnamefont{Macpherson}},
  \bibinfo{author}{\bibfnamefont{D.~N.} \bibnamefont{Johnstone}},
  \bibinfo{author}{\bibfnamefont{V.}~\bibnamefont{Pareek}},
  \bibinfo{author}{\bibfnamefont{E.~M.} \bibnamefont{Tennyson}},
  \bibinfo{author}{\bibfnamefont{S.}~\bibnamefont{Kosar}},
  \bibinfo{author}{\bibfnamefont{F.~U.} \bibnamefont{Kosasih}},
  \bibinfo{author}{\bibfnamefont{M.}~\bibnamefont{Anaya}},
  \bibinfo{author}{\bibfnamefont{M.}~\bibnamefont{Abdi-Jalebi}},
  \bibnamefont{et~al.}, \bibinfo{journal}{Nature}
  \textbf{\bibinfo{volume}{580}}, \bibinfo{pages}{360} (\bibinfo{year}{2020}),
  ISSN \bibinfo{issn}{14764687},
  \urlprefix\url{https://doi.org/10.1038/s41586-020-2184-1}.

\bibitem[{\citenamefont{Ono et~al.}(2020)\citenamefont{Ono, Liu, and
  Qi}}]{Ono2020}
\bibinfo{author}{\bibfnamefont{L.~K.} \bibnamefont{Ono}},
  \bibinfo{author}{\bibfnamefont{S.~F.} \bibnamefont{Liu}}, \bibnamefont{and}
  \bibinfo{author}{\bibfnamefont{Y.}~\bibnamefont{Qi}},
  \bibinfo{journal}{Angew. Chemie Int. Ed.} \textbf{\bibinfo{volume}{59}},
  \bibinfo{pages}{6676} (\bibinfo{year}{2020}), ISSN \bibinfo{issn}{1433-7851},
  \urlprefix\url{https://onlinelibrary.wiley.com/doi/abs/10.1002/anie.201905521}.

\bibitem[{\citenamefont{Heo et~al.}(2017)\citenamefont{Heo, Seo, Lee, Lee,
  Seol, Lee, Park, Kim, Yun, Kim et~al.}}]{Heo2017}
\bibinfo{author}{\bibfnamefont{S.}~\bibnamefont{Heo}},
  \bibinfo{author}{\bibfnamefont{G.}~\bibnamefont{Seo}},
  \bibinfo{author}{\bibfnamefont{Y.}~\bibnamefont{Lee}},
  \bibinfo{author}{\bibfnamefont{D.}~\bibnamefont{Lee}},
  \bibinfo{author}{\bibfnamefont{M.}~\bibnamefont{Seol}},
  \bibinfo{author}{\bibfnamefont{J.}~\bibnamefont{Lee}},
  \bibinfo{author}{\bibfnamefont{J.~B.} \bibnamefont{Park}},
  \bibinfo{author}{\bibfnamefont{K.}~\bibnamefont{Kim}},
  \bibinfo{author}{\bibfnamefont{D.~J.} \bibnamefont{Yun}},
  \bibinfo{author}{\bibfnamefont{Y.~S.} \bibnamefont{Kim}},
  \bibnamefont{et~al.}, \bibinfo{journal}{Energy Environ. Sci.}
  \textbf{\bibinfo{volume}{10}}, \bibinfo{pages}{1128} (\bibinfo{year}{2017}),
  ISSN \bibinfo{issn}{17545706},
  \urlprefix\url{https://pubs.rsc.org/en/content/articlehtml/2017/ee/c7ee00303j}.

\bibitem[{\citenamefont{Blancon et~al.}(2016)\citenamefont{Blancon, Nie,
  Neukirch, Gupta, Tretiak, Cognet, Mohite, and Crochet}}]{Blancon2016}
\bibinfo{author}{\bibfnamefont{J.-C.} \bibnamefont{Blancon}},
  \bibinfo{author}{\bibfnamefont{W.}~\bibnamefont{Nie}},
  \bibinfo{author}{\bibfnamefont{A.~J.} \bibnamefont{Neukirch}},
  \bibinfo{author}{\bibfnamefont{G.}~\bibnamefont{Gupta}},
  \bibinfo{author}{\bibfnamefont{S.}~\bibnamefont{Tretiak}},
  \bibinfo{author}{\bibfnamefont{L.}~\bibnamefont{Cognet}},
  \bibinfo{author}{\bibfnamefont{A.~D.} \bibnamefont{Mohite}},
  \bibnamefont{and} \bibinfo{author}{\bibfnamefont{J.~J.}
  \bibnamefont{Crochet}}, \bibinfo{journal}{Adv. Funct. Mater.}
  \textbf{\bibinfo{volume}{26}}, \bibinfo{pages}{4283} (\bibinfo{year}{2016}),
  ISSN \bibinfo{issn}{1616-301X},
  \urlprefix\url{https://doi.org/10.1002/adfm.201505324}.

\bibitem[{\citenamefont{Leijtens et~al.}(2016)\citenamefont{Leijtens, Eperon,
  Barker, Grancini, Zhang, Ball, Kandada, Snaith, and Petrozza}}]{Leijtens2016}
\bibinfo{author}{\bibfnamefont{T.}~\bibnamefont{Leijtens}},
  \bibinfo{author}{\bibfnamefont{G.~E.} \bibnamefont{Eperon}},
  \bibinfo{author}{\bibfnamefont{A.~J.} \bibnamefont{Barker}},
  \bibinfo{author}{\bibfnamefont{G.}~\bibnamefont{Grancini}},
  \bibinfo{author}{\bibfnamefont{W.}~\bibnamefont{Zhang}},
  \bibinfo{author}{\bibfnamefont{J.~M.} \bibnamefont{Ball}},
  \bibinfo{author}{\bibfnamefont{A.~R.~S.} \bibnamefont{Kandada}},
  \bibinfo{author}{\bibfnamefont{H.~J.} \bibnamefont{Snaith}},
  \bibnamefont{and} \bibinfo{author}{\bibfnamefont{A.}~\bibnamefont{Petrozza}},
  \bibinfo{journal}{Energy Environ. Sci.} \textbf{\bibinfo{volume}{9}},
  \bibinfo{pages}{3472} (\bibinfo{year}{2016}),
  \urlprefix\url{http://dx.doi.org/10.1039/C6EE01729K}.

\bibitem[{\citenamefont{Shi et~al.}(2015{\natexlab{a}})\citenamefont{Shi,
  Adinolfi, Comin, Yuan, Alarousu, Buin, Chen, Hoogland, Rothenberger, Katsiev
  et~al.}}]{Shi2015}
\bibinfo{author}{\bibfnamefont{D.}~\bibnamefont{Shi}},
  \bibinfo{author}{\bibfnamefont{V.}~\bibnamefont{Adinolfi}},
  \bibinfo{author}{\bibfnamefont{R.}~\bibnamefont{Comin}},
  \bibinfo{author}{\bibfnamefont{M.}~\bibnamefont{Yuan}},
  \bibinfo{author}{\bibfnamefont{E.}~\bibnamefont{Alarousu}},
  \bibinfo{author}{\bibfnamefont{A.}~\bibnamefont{Buin}},
  \bibinfo{author}{\bibfnamefont{Y.}~\bibnamefont{Chen}},
  \bibinfo{author}{\bibfnamefont{S.}~\bibnamefont{Hoogland}},
  \bibinfo{author}{\bibfnamefont{A.}~\bibnamefont{Rothenberger}},
  \bibinfo{author}{\bibfnamefont{K.}~\bibnamefont{Katsiev}},
  \bibnamefont{et~al.}, \bibinfo{journal}{Science}
  \textbf{\bibinfo{volume}{347}}, \bibinfo{pages}{519}
  (\bibinfo{year}{2015}{\natexlab{a}}), ISSN \bibinfo{issn}{10959203},
  \urlprefix\url{https://www.science.org/doi/10.1126/science.aaa2725}.

\bibitem[{\citenamefont{de~Quilettes et~al.}(2015)\citenamefont{de~Quilettes,
  Vorpahl, Stranks, Nagaoka, Eperon, Ziffer, Snaith, and
  Ginger}}]{DeQuilettes2015}
\bibinfo{author}{\bibfnamefont{D.~W.} \bibnamefont{de~Quilettes}},
  \bibinfo{author}{\bibfnamefont{S.~M.} \bibnamefont{Vorpahl}},
  \bibinfo{author}{\bibfnamefont{S.~D.} \bibnamefont{Stranks}},
  \bibinfo{author}{\bibfnamefont{H.}~\bibnamefont{Nagaoka}},
  \bibinfo{author}{\bibfnamefont{G.~E.} \bibnamefont{Eperon}},
  \bibinfo{author}{\bibfnamefont{M.~E.} \bibnamefont{Ziffer}},
  \bibinfo{author}{\bibfnamefont{H.~J.} \bibnamefont{Snaith}},
  \bibnamefont{and} \bibinfo{author}{\bibfnamefont{D.~S.}
  \bibnamefont{Ginger}}, \bibinfo{journal}{Science}
  \textbf{\bibinfo{volume}{348}}, \bibinfo{pages}{683 LP }
  (\bibinfo{year}{2015}),
  \urlprefix\url{http://science.sciencemag.org/content/348/6235/683.abstract}.

\bibitem[{\citenamefont{Chen and Zhou}(2020)}]{Chen2020}
\bibinfo{author}{\bibfnamefont{Y.}~\bibnamefont{Chen}} \bibnamefont{and}
  \bibinfo{author}{\bibfnamefont{H.}~\bibnamefont{Zhou}}, \bibinfo{journal}{J.
  Appl. Phys.} \textbf{\bibinfo{volume}{128}}, \bibinfo{pages}{60903}
  (\bibinfo{year}{2020}), ISSN \bibinfo{issn}{0021-8979},
  \urlprefix\url{https://doi.org/10.1063/5.0012384}.

\bibitem[{\citenamefont{Yuan and Huang}(2016)}]{Yuan2016}
\bibinfo{author}{\bibfnamefont{Y.}~\bibnamefont{Yuan}} \bibnamefont{and}
  \bibinfo{author}{\bibfnamefont{J.}~\bibnamefont{Huang}},
  \bibinfo{journal}{Acc. Chem. Res.} \textbf{\bibinfo{volume}{49}},
  \bibinfo{pages}{286} (\bibinfo{year}{2016}), ISSN \bibinfo{issn}{15204898},
  \urlprefix\url{https://pubs.acs.org/doi/abs/10.1021/acs.accounts.5b00420}.

\bibitem[{\citenamefont{Ball and Petrozza}(2016)}]{Ball2016}
\bibinfo{author}{\bibfnamefont{J.~M.} \bibnamefont{Ball}} \bibnamefont{and}
  \bibinfo{author}{\bibfnamefont{A.}~\bibnamefont{Petrozza}},
  \bibinfo{journal}{Nat. Energy} \textbf{\bibinfo{volume}{1}},
  \bibinfo{pages}{16149} (\bibinfo{year}{2016}),
  \urlprefix\url{https://doi.org/10.1038/nenergy.2016.149}.

\bibitem[{\citenamefont{Chen et~al.}(2016)\citenamefont{Chen, Yi, Wu,
  Haroldson, Gartstein, Rodionov, Tikhonov, Zakhidov, Zhu, and
  Podzorov}}]{Chen2016a}
\bibinfo{author}{\bibfnamefont{Y.}~\bibnamefont{Chen}},
  \bibinfo{author}{\bibfnamefont{H.~T.} \bibnamefont{Yi}},
  \bibinfo{author}{\bibfnamefont{X.}~\bibnamefont{Wu}},
  \bibinfo{author}{\bibfnamefont{R.}~\bibnamefont{Haroldson}},
  \bibinfo{author}{\bibfnamefont{Y.~N.} \bibnamefont{Gartstein}},
  \bibinfo{author}{\bibfnamefont{Y.~I.} \bibnamefont{Rodionov}},
  \bibinfo{author}{\bibfnamefont{K.~S.} \bibnamefont{Tikhonov}},
  \bibinfo{author}{\bibfnamefont{A.}~\bibnamefont{Zakhidov}},
  \bibinfo{author}{\bibfnamefont{X.~Y.} \bibnamefont{Zhu}}, \bibnamefont{and}
  \bibinfo{author}{\bibfnamefont{V.}~\bibnamefont{Podzorov}},
  \bibinfo{journal}{Nat. Commun.} \textbf{\bibinfo{volume}{7}},
  \bibinfo{pages}{12253} (\bibinfo{year}{2016}), ISSN
  \bibinfo{issn}{2041-1723},
  \urlprefix\url{https://doi.org/10.1038/ncomms12253}.

\bibitem[{\citenamefont{Adinolfi et~al.}(2016)\citenamefont{Adinolfi, Yuan,
  Comin, Thibau, Shi, Saidaminov, Kanjanaboos, Kopilovic, Hoogland, Lu
  et~al.}}]{Adinolfi2016}
\bibinfo{author}{\bibfnamefont{V.}~\bibnamefont{Adinolfi}},
  \bibinfo{author}{\bibfnamefont{M.}~\bibnamefont{Yuan}},
  \bibinfo{author}{\bibfnamefont{R.}~\bibnamefont{Comin}},
  \bibinfo{author}{\bibfnamefont{E.~S.} \bibnamefont{Thibau}},
  \bibinfo{author}{\bibfnamefont{D.}~\bibnamefont{Shi}},
  \bibinfo{author}{\bibfnamefont{M.~I.} \bibnamefont{Saidaminov}},
  \bibinfo{author}{\bibfnamefont{P.}~\bibnamefont{Kanjanaboos}},
  \bibinfo{author}{\bibfnamefont{D.}~\bibnamefont{Kopilovic}},
  \bibinfo{author}{\bibfnamefont{S.}~\bibnamefont{Hoogland}},
  \bibinfo{author}{\bibfnamefont{Z.-H.} \bibnamefont{Lu}},
  \bibnamefont{et~al.}, \bibinfo{journal}{Adv. Mater.}
  \textbf{\bibinfo{volume}{28}}, \bibinfo{pages}{3406} (\bibinfo{year}{2016}),
  \urlprefix\url{https://onlinelibrary.wiley.com/doi/abs/10.1002/adma.201505162}.

\bibitem[{\citenamefont{Samiee et~al.}(2014)\citenamefont{Samiee, Konduri,
  Ganapathy, Kottokkaran, Abbas, Kitahara, Joshi, Zhang, Noack, and
  Dalal}}]{Samiee2014a}
\bibinfo{author}{\bibfnamefont{M.}~\bibnamefont{Samiee}},
  \bibinfo{author}{\bibfnamefont{S.}~\bibnamefont{Konduri}},
  \bibinfo{author}{\bibfnamefont{B.}~\bibnamefont{Ganapathy}},
  \bibinfo{author}{\bibfnamefont{R.}~\bibnamefont{Kottokkaran}},
  \bibinfo{author}{\bibfnamefont{H.~A.} \bibnamefont{Abbas}},
  \bibinfo{author}{\bibfnamefont{A.}~\bibnamefont{Kitahara}},
  \bibinfo{author}{\bibfnamefont{P.}~\bibnamefont{Joshi}},
  \bibinfo{author}{\bibfnamefont{L.}~\bibnamefont{Zhang}},
  \bibinfo{author}{\bibfnamefont{M.}~\bibnamefont{Noack}}, \bibnamefont{and}
  \bibinfo{author}{\bibfnamefont{V.}~\bibnamefont{Dalal}},
  \bibinfo{journal}{Appl. Phys. Lett.} \textbf{\bibinfo{volume}{105}},
  \bibinfo{pages}{153502} (\bibinfo{year}{2014}), ISSN
  \bibinfo{issn}{0003-6951}, \urlprefix\url{https://doi.org/10.1063/1.4897329}.

\bibitem[{\citenamefont{Qin et~al.}(2020)\citenamefont{Qin, Xue, Zhang, Hu,
  Liu, Li, Qin, Ma, Zhu, Yan et~al.}}]{Qin2020}
\bibinfo{author}{\bibfnamefont{M.}~\bibnamefont{Qin}},
  \bibinfo{author}{\bibfnamefont{H.}~\bibnamefont{Xue}},
  \bibinfo{author}{\bibfnamefont{H.}~\bibnamefont{Zhang}},
  \bibinfo{author}{\bibfnamefont{H.}~\bibnamefont{Hu}},
  \bibinfo{author}{\bibfnamefont{K.}~\bibnamefont{Liu}},
  \bibinfo{author}{\bibfnamefont{Y.}~\bibnamefont{Li}},
  \bibinfo{author}{\bibfnamefont{Z.}~\bibnamefont{Qin}},
  \bibinfo{author}{\bibfnamefont{J.}~\bibnamefont{Ma}},
  \bibinfo{author}{\bibfnamefont{H.}~\bibnamefont{Zhu}},
  \bibinfo{author}{\bibfnamefont{K.}~\bibnamefont{Yan}}, \bibnamefont{et~al.},
  \bibinfo{journal}{Adv. Mater.} \textbf{\bibinfo{volume}{32}},
  \bibinfo{pages}{2004630} (\bibinfo{year}{2020}), ISSN
  \bibinfo{issn}{0935-9648},
  \urlprefix\url{https://onlinelibrary.wiley.com/doi/10.1002/adma.202004630}.

\bibitem[{\citenamefont{Li et~al.}(2020)\citenamefont{Li, Luo, Chen, Niu,
  Zhang, Lu, Kumar, Jiang, Liu, Guo et~al.}}]{Li2020}
\bibinfo{author}{\bibfnamefont{N.}~\bibnamefont{Li}},
  \bibinfo{author}{\bibfnamefont{Y.}~\bibnamefont{Luo}},
  \bibinfo{author}{\bibfnamefont{Z.}~\bibnamefont{Chen}},
  \bibinfo{author}{\bibfnamefont{X.}~\bibnamefont{Niu}},
  \bibinfo{author}{\bibfnamefont{X.}~\bibnamefont{Zhang}},
  \bibinfo{author}{\bibfnamefont{J.}~\bibnamefont{Lu}},
  \bibinfo{author}{\bibfnamefont{R.}~\bibnamefont{Kumar}},
  \bibinfo{author}{\bibfnamefont{J.}~\bibnamefont{Jiang}},
  \bibinfo{author}{\bibfnamefont{H.}~\bibnamefont{Liu}},
  \bibinfo{author}{\bibfnamefont{X.}~\bibnamefont{Guo}}, \bibnamefont{et~al.},
  \bibinfo{journal}{Joule} \textbf{\bibinfo{volume}{4}}, \bibinfo{pages}{1743}
  (\bibinfo{year}{2020}), ISSN \bibinfo{issn}{25424351},
  \urlprefix\url{https://linkinghub.elsevier.com/retrieve/pii/S2542435120302427}.

\bibitem[{\citenamefont{Zhou et~al.}(2019)\citenamefont{Zhou, Xue, Jia, Brocks,
  Tao, and Zhao}}]{Zhou2019}
\bibinfo{author}{\bibfnamefont{Y.}~\bibnamefont{Zhou}},
  \bibinfo{author}{\bibfnamefont{H.}~\bibnamefont{Xue}},
  \bibinfo{author}{\bibfnamefont{Y.}~\bibnamefont{Jia}},
  \bibinfo{author}{\bibfnamefont{G.}~\bibnamefont{Brocks}},
  \bibinfo{author}{\bibfnamefont{S.}~\bibnamefont{Tao}}, \bibnamefont{and}
  \bibinfo{author}{\bibfnamefont{N.}~\bibnamefont{Zhao}},
  \bibinfo{journal}{Adv. Funct. Mater.} \textbf{\bibinfo{volume}{29}},
  \bibinfo{pages}{1905739} (\bibinfo{year}{2019}), ISSN
  \bibinfo{issn}{1616-301X},
  \urlprefix\url{https://onlinelibrary.wiley.com/doi/abs/10.1002/adfm.201905739}.

\bibitem[{\citenamefont{Yang et~al.}(2017)\citenamefont{Yang, Park, Jung, Jeon,
  Kim, Lee, Shin, Seo, Kim, Noh et~al.}}]{Yang2017}
\bibinfo{author}{\bibfnamefont{W.~S.} \bibnamefont{Yang}},
  \bibinfo{author}{\bibfnamefont{B.~W.} \bibnamefont{Park}},
  \bibinfo{author}{\bibfnamefont{E.~H.} \bibnamefont{Jung}},
  \bibinfo{author}{\bibfnamefont{N.~J.} \bibnamefont{Jeon}},
  \bibinfo{author}{\bibfnamefont{Y.~C.} \bibnamefont{Kim}},
  \bibinfo{author}{\bibfnamefont{D.~U.} \bibnamefont{Lee}},
  \bibinfo{author}{\bibfnamefont{S.~S.} \bibnamefont{Shin}},
  \bibinfo{author}{\bibfnamefont{J.}~\bibnamefont{Seo}},
  \bibinfo{author}{\bibfnamefont{E.~K.} \bibnamefont{Kim}},
  \bibinfo{author}{\bibfnamefont{J.~H.} \bibnamefont{Noh}},
  \bibnamefont{et~al.}, \bibinfo{journal}{Science}
  \textbf{\bibinfo{volume}{356}}, \bibinfo{pages}{1376} (\bibinfo{year}{2017}),
  \urlprefix\url{https://www.science.org/doi/10.1126/science.aan2301}.

\bibitem[{\citenamefont{Saliba et~al.}(2016)\citenamefont{Saliba, Matsui,
  Domanski, Seo, Ummadisingu, Zakeeruddin, Correa-Baena, Tress, Abate, Hagfeldt
  et~al.}}]{Saliba2016}
\bibinfo{author}{\bibfnamefont{M.}~\bibnamefont{Saliba}},
  \bibinfo{author}{\bibfnamefont{T.}~\bibnamefont{Matsui}},
  \bibinfo{author}{\bibfnamefont{K.}~\bibnamefont{Domanski}},
  \bibinfo{author}{\bibfnamefont{J.~Y.} \bibnamefont{Seo}},
  \bibinfo{author}{\bibfnamefont{A.}~\bibnamefont{Ummadisingu}},
  \bibinfo{author}{\bibfnamefont{S.~M.} \bibnamefont{Zakeeruddin}},
  \bibinfo{author}{\bibfnamefont{J.~P.} \bibnamefont{Correa-Baena}},
  \bibinfo{author}{\bibfnamefont{W.~R.} \bibnamefont{Tress}},
  \bibinfo{author}{\bibfnamefont{A.}~\bibnamefont{Abate}},
  \bibinfo{author}{\bibfnamefont{A.}~\bibnamefont{Hagfeldt}},
  \bibnamefont{et~al.}, \bibinfo{journal}{Science}
  \textbf{\bibinfo{volume}{354}}, \bibinfo{pages}{206} (\bibinfo{year}{2016}),
  ISSN \bibinfo{issn}{10959203},
  \urlprefix\url{https://www.science.org/doi/10.1126/science.aah5557}.

\bibitem[{\citenamefont{Hieulle et~al.}(2018)\citenamefont{Hieulle, Stecker,
  Ohmann, Ono, and Qi}}]{Hieulle2018}
\bibinfo{author}{\bibfnamefont{J.}~\bibnamefont{Hieulle}},
  \bibinfo{author}{\bibfnamefont{C.}~\bibnamefont{Stecker}},
  \bibinfo{author}{\bibfnamefont{R.}~\bibnamefont{Ohmann}},
  \bibinfo{author}{\bibfnamefont{L.~K.} \bibnamefont{Ono}}, \bibnamefont{and}
  \bibinfo{author}{\bibfnamefont{Y.}~\bibnamefont{Qi}}, \bibinfo{journal}{Small
  Methods} \textbf{\bibinfo{volume}{2}}, \bibinfo{pages}{1700295}
  (\bibinfo{year}{2018}), ISSN \bibinfo{issn}{23669608},
  \urlprefix\url{http://doi.wiley.com/10.1002/smtd.201700295}.

\bibitem[{\citenamefont{Shih et~al.}(2017)\citenamefont{Shih, Li, Hsieh, Wang,
  Yang, Chiu, Chang, and Chen}}]{Shih2017}
\bibinfo{author}{\bibfnamefont{M.~C.} \bibnamefont{Shih}},
  \bibinfo{author}{\bibfnamefont{S.~S.} \bibnamefont{Li}},
  \bibinfo{author}{\bibfnamefont{C.~H.} \bibnamefont{Hsieh}},
  \bibinfo{author}{\bibfnamefont{Y.~C.} \bibnamefont{Wang}},
  \bibinfo{author}{\bibfnamefont{H.~D.} \bibnamefont{Yang}},
  \bibinfo{author}{\bibfnamefont{Y.~P.} \bibnamefont{Chiu}},
  \bibinfo{author}{\bibfnamefont{C.~S.} \bibnamefont{Chang}}, \bibnamefont{and}
  \bibinfo{author}{\bibfnamefont{C.~W.} \bibnamefont{Chen}},
  \bibinfo{journal}{Nano Lett.} \textbf{\bibinfo{volume}{17}},
  \bibinfo{pages}{1154} (\bibinfo{year}{2017}), ISSN \bibinfo{issn}{15306992},
  \urlprefix\url{https://pubs.acs.org/doi/10.1021/acs.nanolett.6b04803}.

\bibitem[{\citenamefont{Ohmann et~al.}(2015)\citenamefont{Ohmann, Ono, Kim,
  Lin, Lee, Li, Park, and Qi}}]{Ohmann2015}
\bibinfo{author}{\bibfnamefont{R.}~\bibnamefont{Ohmann}},
  \bibinfo{author}{\bibfnamefont{L.~K.} \bibnamefont{Ono}},
  \bibinfo{author}{\bibfnamefont{H.~S.} \bibnamefont{Kim}},
  \bibinfo{author}{\bibfnamefont{H.}~\bibnamefont{Lin}},
  \bibinfo{author}{\bibfnamefont{M.~V.} \bibnamefont{Lee}},
  \bibinfo{author}{\bibfnamefont{Y.}~\bibnamefont{Li}},
  \bibinfo{author}{\bibfnamefont{N.~G.} \bibnamefont{Park}}, \bibnamefont{and}
  \bibinfo{author}{\bibfnamefont{Y.}~\bibnamefont{Qi}}, \bibinfo{journal}{J.
  Am. Chem. Soc.} \textbf{\bibinfo{volume}{137}}, \bibinfo{pages}{16049}
  (\bibinfo{year}{2015}), ISSN \bibinfo{issn}{15205126},
  \urlprefix\url{https://pubs.acs.org/doi/10.1021/jacs.5b08227}.

\bibitem[{\citenamefont{Yin et~al.}(2014)\citenamefont{Yin, Shi, and
  Yan}}]{Yin2014}
\bibinfo{author}{\bibfnamefont{W.-J.} \bibnamefont{Yin}},
  \bibinfo{author}{\bibfnamefont{T.}~\bibnamefont{Shi}}, \bibnamefont{and}
  \bibinfo{author}{\bibfnamefont{Y.}~\bibnamefont{Yan}},
  \bibinfo{journal}{Appl. Phys. Lett.} \textbf{\bibinfo{volume}{104}},
  \bibinfo{pages}{63903} (\bibinfo{year}{2014}),
  \urlprefix\url{https://aip.scitation.org/doi/abs/10.1063/1.4864778}.

\bibitem[{\citenamefont{Meggiolaro and {De
  Angelis}}(2018)}]{Meggiolaro2018review}
\bibinfo{author}{\bibfnamefont{D.}~\bibnamefont{Meggiolaro}} \bibnamefont{and}
  \bibinfo{author}{\bibfnamefont{F.}~\bibnamefont{{De Angelis}}},
  \bibinfo{journal}{ACS Energy Lett.} \textbf{\bibinfo{volume}{3}},
  \bibinfo{pages}{2206} (\bibinfo{year}{2018}),
  \urlprefix\url{https://doi.org/10.1021/acsenergylett.8b01212}.

\bibitem[{\citenamefont{Meggiolaro et~al.}(2018)\citenamefont{Meggiolaro,
  Motti, Mosconi, Barker, Ball, {Andrea Riccardo Perini}, Deschler, Petrozza,
  and {De Angelis}}}]{Daniele2018iodine}
\bibinfo{author}{\bibfnamefont{D.}~\bibnamefont{Meggiolaro}},
  \bibinfo{author}{\bibfnamefont{S.~G.} \bibnamefont{Motti}},
  \bibinfo{author}{\bibfnamefont{E.}~\bibnamefont{Mosconi}},
  \bibinfo{author}{\bibfnamefont{A.~J.} \bibnamefont{Barker}},
  \bibinfo{author}{\bibfnamefont{J.}~\bibnamefont{Ball}},
  \bibinfo{author}{\bibfnamefont{C.}~\bibnamefont{{Andrea Riccardo Perini}}},
  \bibinfo{author}{\bibfnamefont{F.}~\bibnamefont{Deschler}},
  \bibinfo{author}{\bibfnamefont{A.}~\bibnamefont{Petrozza}}, \bibnamefont{and}
  \bibinfo{author}{\bibfnamefont{F.}~\bibnamefont{{De Angelis}}},
  \bibinfo{journal}{Energy Environ. Sci.} \textbf{\bibinfo{volume}{11}},
  \bibinfo{pages}{702} (\bibinfo{year}{2018}), ISSN \bibinfo{issn}{1754-5692},
  \urlprefix\url{http://dx.doi.org/10.1039/C8EE00124C}.

\bibitem[{\citenamefont{Buin et~al.}(2015)\citenamefont{Buin, Comin, Xu, Ip,
  and Sargent}}]{Buin2015}
\bibinfo{author}{\bibfnamefont{A.}~\bibnamefont{Buin}},
  \bibinfo{author}{\bibfnamefont{R.}~\bibnamefont{Comin}},
  \bibinfo{author}{\bibfnamefont{J.}~\bibnamefont{Xu}},
  \bibinfo{author}{\bibfnamefont{A.~H.} \bibnamefont{Ip}}, \bibnamefont{and}
  \bibinfo{author}{\bibfnamefont{E.~H.} \bibnamefont{Sargent}},
  \bibinfo{journal}{Chem. Mater.} \textbf{\bibinfo{volume}{27}},
  \bibinfo{pages}{4405} (\bibinfo{year}{2015}), ISSN \bibinfo{issn}{0897-4756},
  \urlprefix\url{https://doi.org/10.1021/acs.chemmater.5b01909}.

\bibitem[{\citenamefont{Shi et~al.}(2015{\natexlab{b}})\citenamefont{Shi, Yin,
  Hong, Zhu, and Yan}}]{Shi2015a}
\bibinfo{author}{\bibfnamefont{T.}~\bibnamefont{Shi}},
  \bibinfo{author}{\bibfnamefont{W.-J.} \bibnamefont{Yin}},
  \bibinfo{author}{\bibfnamefont{F.}~\bibnamefont{Hong}},
  \bibinfo{author}{\bibfnamefont{K.}~\bibnamefont{Zhu}}, \bibnamefont{and}
  \bibinfo{author}{\bibfnamefont{Y.}~\bibnamefont{Yan}},
  \bibinfo{journal}{Appl. Phys. Lett.} \textbf{\bibinfo{volume}{106}},
  \bibinfo{pages}{103902} (\bibinfo{year}{2015}{\natexlab{b}}),
  \urlprefix\url{https://doi.org/10.1063/1.4914544}.

\bibitem[{\citenamefont{Meggiolaro et~al.}(2020)\citenamefont{Meggiolaro,
  Ricciarelli, Alasmari, Alasmary, and {De Angelis}}}]{Meggiolaro2020a}
\bibinfo{author}{\bibfnamefont{D.}~\bibnamefont{Meggiolaro}},
  \bibinfo{author}{\bibfnamefont{D.}~\bibnamefont{Ricciarelli}},
  \bibinfo{author}{\bibfnamefont{A.~A.} \bibnamefont{Alasmari}},
  \bibinfo{author}{\bibfnamefont{F.~A.~S.} \bibnamefont{Alasmary}},
  \bibnamefont{and} \bibinfo{author}{\bibfnamefont{F.}~\bibnamefont{{De
  Angelis}}}, \bibinfo{journal}{J. Phys. Chem. Lett.}
  \textbf{\bibinfo{volume}{11}}, \bibinfo{pages}{3546} (\bibinfo{year}{2020}),
  ISSN \bibinfo{issn}{19487185},
  \urlprefix\url{https://dx.doi.org/10.1021/acs.jpclett.0c00725}.

\bibitem[{\citenamefont{Shi et~al.}(2017)\citenamefont{Shi, Zhang, Meng, Teng,
  Liu, Yang, Yan, Yip, and Zhao}}]{Shi2017}
\bibinfo{author}{\bibfnamefont{T.}~\bibnamefont{Shi}},
  \bibinfo{author}{\bibfnamefont{H.-S.} \bibnamefont{Zhang}},
  \bibinfo{author}{\bibfnamefont{W.}~\bibnamefont{Meng}},
  \bibinfo{author}{\bibfnamefont{Q.}~\bibnamefont{Teng}},
  \bibinfo{author}{\bibfnamefont{M.}~\bibnamefont{Liu}},
  \bibinfo{author}{\bibfnamefont{X.}~\bibnamefont{Yang}},
  \bibinfo{author}{\bibfnamefont{Y.}~\bibnamefont{Yan}},
  \bibinfo{author}{\bibfnamefont{H.-L.} \bibnamefont{Yip}}, \bibnamefont{and}
  \bibinfo{author}{\bibfnamefont{Y.-J.} \bibnamefont{Zhao}},
  \bibinfo{journal}{J. Mater. Chem. A} \textbf{\bibinfo{volume}{5}},
  \bibinfo{pages}{15124} (\bibinfo{year}{2017}), ISSN
  \bibinfo{issn}{2050-7488},
  \urlprefix\url{http://dx.doi.org/10.1039/C7TA02662E}.

\bibitem[{\citenamefont{Liu and Yam}(2018)}]{Liu2018}
\bibinfo{author}{\bibfnamefont{N.}~\bibnamefont{Liu}} \bibnamefont{and}
  \bibinfo{author}{\bibfnamefont{C.}~\bibnamefont{Yam}},
  \bibinfo{journal}{Phys. Chem. Chem. Phys.} \textbf{\bibinfo{volume}{20}},
  \bibinfo{pages}{6800} (\bibinfo{year}{2018}), ISSN \bibinfo{issn}{1463-9076},
  \urlprefix\url{http://dx.doi.org/10.1039/C8CP00280K}.

\bibitem[{\citenamefont{Huang et~al.}(2018)\citenamefont{Huang, Yin, and
  He}}]{Huang2018}
\bibinfo{author}{\bibfnamefont{Y.}~\bibnamefont{Huang}},
  \bibinfo{author}{\bibfnamefont{W.-J.} \bibnamefont{Yin}}, \bibnamefont{and}
  \bibinfo{author}{\bibfnamefont{Y.}~\bibnamefont{He}}, \bibinfo{journal}{J.
  Phys. Chem. C} \textbf{\bibinfo{volume}{122}}, \bibinfo{pages}{1345}
  (\bibinfo{year}{2018}), ISSN \bibinfo{issn}{1932-7447},
  \urlprefix\url{https://doi.org/10.1021/acs.jpcc.7b10045}.

\bibitem[{\citenamefont{Kang and Wang}(2017)}]{Kang2017}
\bibinfo{author}{\bibfnamefont{J.}~\bibnamefont{Kang}} \bibnamefont{and}
  \bibinfo{author}{\bibfnamefont{L.-W.} \bibnamefont{Wang}},
  \bibinfo{journal}{J. Phys. Chem. Lett.} \textbf{\bibinfo{volume}{8}},
  \bibinfo{pages}{489} (\bibinfo{year}{2017}), ISSN \bibinfo{issn}{1948-7185},
  \urlprefix\url{https://doi.org/10.1021/acs.jpclett.6b02800}.

\bibitem[{\citenamefont{Xu et~al.}(2014)\citenamefont{Xu, Chen, Xiang, Gong,
  and Wei}}]{Xu2014}
\bibinfo{author}{\bibfnamefont{P.}~\bibnamefont{Xu}},
  \bibinfo{author}{\bibfnamefont{S.}~\bibnamefont{Chen}},
  \bibinfo{author}{\bibfnamefont{H.-J.} \bibnamefont{Xiang}},
  \bibinfo{author}{\bibfnamefont{X.-G.} \bibnamefont{Gong}}, \bibnamefont{and}
  \bibinfo{author}{\bibfnamefont{S.-H.} \bibnamefont{Wei}},
  \bibinfo{journal}{Chem. Mater.} \textbf{\bibinfo{volume}{26}},
  \bibinfo{pages}{6068} (\bibinfo{year}{2014}), ISSN \bibinfo{issn}{0897-4756},
  \urlprefix\url{https://doi.org/10.1021/cm503122j}.

\bibitem[{\citenamefont{Liang et~al.}(2021)\citenamefont{Liang, Cui, Li,
  Stampfl, Ringer, and Zheng}}]{Liang2021}
\bibinfo{author}{\bibfnamefont{Y.}~\bibnamefont{Liang}},
  \bibinfo{author}{\bibfnamefont{X.}~\bibnamefont{Cui}},
  \bibinfo{author}{\bibfnamefont{F.}~\bibnamefont{Li}},
  \bibinfo{author}{\bibfnamefont{C.}~\bibnamefont{Stampfl}},
  \bibinfo{author}{\bibfnamefont{S.~P.} \bibnamefont{Ringer}},
  \bibnamefont{and} \bibinfo{author}{\bibfnamefont{R.}~\bibnamefont{Zheng}},
  \bibinfo{journal}{Phys. Rev. Mater.} \textbf{\bibinfo{volume}{5}},
  \bibinfo{pages}{35405} (\bibinfo{year}{2021}), ISSN
  \bibinfo{issn}{2475-9953},
  \urlprefix\url{https://link.aps.org/doi/10.1103/PhysRevMaterials.5.035405}.

\bibitem[{\citenamefont{Du}(2015)}]{Du2015}
\bibinfo{author}{\bibfnamefont{M.-H.} \bibnamefont{Du}}, \bibinfo{journal}{J.
  Phys. Chem. Lett.} \textbf{\bibinfo{volume}{6}}, \bibinfo{pages}{1461}
  (\bibinfo{year}{2015}), ISSN \bibinfo{issn}{1948-7185},
  \urlprefix\url{https://doi.org/10.1021/acs.jpclett.5b00199}.

\bibitem[{\citenamefont{Agiorgousis et~al.}(2014)\citenamefont{Agiorgousis,
  Sun, Zeng, and Zhang}}]{Michael2014}
\bibinfo{author}{\bibfnamefont{M.~L.} \bibnamefont{Agiorgousis}},
  \bibinfo{author}{\bibfnamefont{Y.-Y.} \bibnamefont{Sun}},
  \bibinfo{author}{\bibfnamefont{H.}~\bibnamefont{Zeng}}, \bibnamefont{and}
  \bibinfo{author}{\bibfnamefont{S.}~\bibnamefont{Zhang}}, \bibinfo{journal}{J.
  Am. Chem. Soc.} \textbf{\bibinfo{volume}{136}}, \bibinfo{pages}{14570}
  (\bibinfo{year}{2014}), ISSN \bibinfo{issn}{0002-7863},
  \urlprefix\url{https://doi.org/10.1021/ja5079305}.

\bibitem[{\citenamefont{Xue et~al.}(2021)\citenamefont{Xue, Brocks, and
  Tao}}]{Xue2021}
\bibinfo{author}{\bibfnamefont{H.}~\bibnamefont{Xue}},
  \bibinfo{author}{\bibfnamefont{G.}~\bibnamefont{Brocks}}, \bibnamefont{and}
  \bibinfo{author}{\bibfnamefont{S.}~\bibnamefont{Tao}},
  \bibinfo{journal}{Phys. Rev. Mater.} \textbf{\bibinfo{volume}{5}},
  \bibinfo{pages}{125408} (\bibinfo{year}{2021}), ISSN
  \bibinfo{issn}{2475-9953},
  \urlprefix\url{https://link.aps.org/doi/10.1103/PhysRevMaterials.5.125408}.

\bibitem[{\citenamefont{Kresse and Hafner}(1993)}]{Kresse1993}
\bibinfo{author}{\bibfnamefont{G.}~\bibnamefont{Kresse}} \bibnamefont{and}
  \bibinfo{author}{\bibfnamefont{J.}~\bibnamefont{Hafner}},
  \bibinfo{journal}{Phys. Rev. B} \textbf{\bibinfo{volume}{47}},
  \bibinfo{pages}{558} (\bibinfo{year}{1993}),
  \urlprefix\url{https://link.aps.org/doi/10.1103/PhysRevB.47.558}.

\bibitem[{\citenamefont{Kresse and
  Furthm{\"{u}}ller}(1996{\natexlab{a}})}]{Kresse1996}
\bibinfo{author}{\bibfnamefont{G.}~\bibnamefont{Kresse}} \bibnamefont{and}
  \bibinfo{author}{\bibfnamefont{J.}~\bibnamefont{Furthm{\"{u}}ller}},
  \bibinfo{journal}{Phys. Rev. B} \textbf{\bibinfo{volume}{54}},
  \bibinfo{pages}{11169} (\bibinfo{year}{1996}{\natexlab{a}}),
  \urlprefix\url{https://link.aps.org/doi/10.1103/PhysRevB.54.11169}.

\bibitem[{\citenamefont{Kresse and
  Furthm{\"{u}}ller}(1996{\natexlab{b}})}]{Kresse1996a}
\bibinfo{author}{\bibfnamefont{G.}~\bibnamefont{Kresse}} \bibnamefont{and}
  \bibinfo{author}{\bibfnamefont{J.}~\bibnamefont{Furthm{\"{u}}ller}},
  \bibinfo{journal}{Comput. Mater. Sci.} \textbf{\bibinfo{volume}{6}},
  \bibinfo{pages}{15} (\bibinfo{year}{1996}{\natexlab{b}}),
  \urlprefix\url{http://www.sciencedirect.com/science/article/pii/0927025696000080}.

\bibitem[{\citenamefont{Peng et~al.}(2016)\citenamefont{Peng, Yang, Perdew, and
  Sun}}]{Peng2016}
\bibinfo{author}{\bibfnamefont{H.}~\bibnamefont{Peng}},
  \bibinfo{author}{\bibfnamefont{Z.-H.} \bibnamefont{Yang}},
  \bibinfo{author}{\bibfnamefont{J.~P.} \bibnamefont{Perdew}},
  \bibnamefont{and} \bibinfo{author}{\bibfnamefont{J.}~\bibnamefont{Sun}},
  \bibinfo{journal}{Phys. Rev. X} \textbf{\bibinfo{volume}{6}},
  \bibinfo{pages}{41005} (\bibinfo{year}{2016}),
  \urlprefix\url{https://link.aps.org/doi/10.1103/PhysRevX.6.041005}.

\bibitem[{\citenamefont{Sun et~al.}(2015)\citenamefont{Sun, Ruzsinszky, and
  Perdew}}]{Sun2015}
\bibinfo{author}{\bibfnamefont{J.}~\bibnamefont{Sun}},
  \bibinfo{author}{\bibfnamefont{A.}~\bibnamefont{Ruzsinszky}},
  \bibnamefont{and} \bibinfo{author}{\bibfnamefont{J.~P.}
  \bibnamefont{Perdew}}, \bibinfo{journal}{Phys. Rev. Lett.}
  \textbf{\bibinfo{volume}{115}}, \bibinfo{pages}{36402}
  (\bibinfo{year}{2015}),
  \urlprefix\url{https://link.aps.org/doi/10.1103/PhysRevLett.115.036402}.

\bibitem[{\citenamefont{Sabatini et~al.}(2013)\citenamefont{Sabatini, Gorni,
  and de~Gironcoli}}]{Sabatini2013}
\bibinfo{author}{\bibfnamefont{R.}~\bibnamefont{Sabatini}},
  \bibinfo{author}{\bibfnamefont{T.}~\bibnamefont{Gorni}}, \bibnamefont{and}
  \bibinfo{author}{\bibfnamefont{S.}~\bibnamefont{de~Gironcoli}},
  \bibinfo{journal}{Phys. Rev. B} \textbf{\bibinfo{volume}{87}},
  \bibinfo{pages}{041108} (\bibinfo{year}{2013}), ISSN
  \bibinfo{issn}{1098-0121},
  \urlprefix\url{https://link.aps.org/doi/10.1103/PhysRevB.87.041108}.

\bibitem[{\citenamefont{Manz and Limas}(2016{\natexlab{a}})}]{Manz2016}
\bibinfo{author}{\bibfnamefont{T.~A.} \bibnamefont{Manz}} \bibnamefont{and}
  \bibinfo{author}{\bibfnamefont{N.~G.} \bibnamefont{Limas}},
  \textbf{\bibinfo{volume}{6}}, \bibinfo{pages}{47771}
  (\bibinfo{year}{2016}{\natexlab{a}}),
  \urlprefix\url{http://dx.doi.org/10.1039/C6RA04656H}.

\bibitem[{\citenamefont{Manz}(2017)}]{Manz2017}
\bibinfo{author}{\bibfnamefont{T.~A.} \bibnamefont{Manz}},
  \bibinfo{journal}{RSC Adv.} \textbf{\bibinfo{volume}{7}},
  \bibinfo{pages}{45552} (\bibinfo{year}{2017}),
  \urlprefix\url{http://dx.doi.org/10.1039/C7RA07400J}.

\bibitem[{\citenamefont{Manz and Limas}(2016{\natexlab{b}})}]{Manz2016program}
\bibinfo{author}{\bibfnamefont{T.~A.} \bibnamefont{Manz}} \bibnamefont{and}
  \bibinfo{author}{\bibfnamefont{N.~G.} \bibnamefont{Limas}},
  \emph{\bibinfo{title}{{Program Computing DDEC Atomic Charges}}}
  (\bibinfo{year}{2016}{\natexlab{b}}),
  \urlprefix\url{https://sourceforge.net/projects/ddec/}.

\bibitem[{\citenamefont{Tao et~al.}(2019)\citenamefont{Tao, Schmidt, Brocks,
  Jiang, Tranca, Meerholz, and Olthof}}]{Tao2019}
\bibinfo{author}{\bibfnamefont{S.}~\bibnamefont{Tao}},
  \bibinfo{author}{\bibfnamefont{I.}~\bibnamefont{Schmidt}},
  \bibinfo{author}{\bibfnamefont{G.}~\bibnamefont{Brocks}},
  \bibinfo{author}{\bibfnamefont{J.}~\bibnamefont{Jiang}},
  \bibinfo{author}{\bibfnamefont{I.}~\bibnamefont{Tranca}},
  \bibinfo{author}{\bibfnamefont{K.}~\bibnamefont{Meerholz}}, \bibnamefont{and}
  \bibinfo{author}{\bibfnamefont{S.}~\bibnamefont{Olthof}},
  \bibinfo{journal}{Nat. Commun.} \textbf{\bibinfo{volume}{10}},
  \bibinfo{pages}{2560} (\bibinfo{year}{2019}),
  \urlprefix\url{https://doi.org/10.1038/s41467-019-10468-7}.

\bibitem[{\citenamefont{Walsh}(2015)}]{walsh_2015}
\bibinfo{author}{\bibfnamefont{A.}~\bibnamefont{Walsh}},
  \emph{\bibinfo{title}{{WMD-group/hybrid-perovskites/2015-ch3nh3pbi3-phonons-PBEsol/CH3NH3PbI3-tetragonal.POSCAR.vasp}}}
  (\bibinfo{year}{2015}),
  \urlprefix\url{https://github.com/WMD-group/hybrid-perovskites/blob/master/2015_ch3nh3pbi3_phonons_PBEsol/CH3NH3PbI3_tetragonal.POSCAR.vasp}.

\bibitem[{\citenamefont{Walsh}(2016)}]{walsh_2016}
\bibinfo{author}{\bibfnamefont{A.}~\bibnamefont{Walsh}},
  \emph{\bibinfo{title}{{WMD-group/hybrid-perovskites/2015-cssni3-phonons-PBEsol/CsSnI3-Bg.POSCAR.vasp}}}
  (\bibinfo{year}{2016}),
  \urlprefix\url{https://github.com/WMD-group/hybrid-perovskites/blob/master/2015_cssni3_phonons_PBEsol/CsSnI3-Bg.POSCAR.vasp}.

\bibitem[{\citenamefont{Walsh}(2017)}]{walsh_2017}
\bibinfo{author}{\bibfnamefont{A.}~\bibnamefont{Walsh}},
  \emph{\bibinfo{title}{{WMD-group/hybrid-perovskites/2015-fapi-PBEsol/FAPbI3-tetragonal-P42mnm.POSCAR.vasp}}}
  (\bibinfo{year}{2017}),
  \urlprefix\url{https://github.com/WMD-group/hybrid-perovskites/blob/master/2015_fapi_PBEsol/FAPbI3_tetragonal_P42mnm.POSCAR.vasp}.

\bibitem[{\citenamefont{de~Walle and Neugebauer}(2004)}]{Walle2004}
\bibinfo{author}{\bibfnamefont{C.~G.~V.} \bibnamefont{de~Walle}}
  \bibnamefont{and}
  \bibinfo{author}{\bibfnamefont{J.}~\bibnamefont{Neugebauer}},
  \bibinfo{journal}{J. Appl. Phys.} \textbf{\bibinfo{volume}{95}},
  \bibinfo{pages}{3851} (\bibinfo{year}{2004}),
  \urlprefix\url{https://aip.scitation.org/doi/abs/10.1063/1.1682673}.

\bibitem[{\citenamefont{Freysoldt et~al.}(2014)\citenamefont{Freysoldt,
  Grabowski, Hickel, Neugebauer, Kresse, Janotti, and {Van de
  Walle}}}]{Freysoldt2014}
\bibinfo{author}{\bibfnamefont{C.}~\bibnamefont{Freysoldt}},
  \bibinfo{author}{\bibfnamefont{B.}~\bibnamefont{Grabowski}},
  \bibinfo{author}{\bibfnamefont{T.}~\bibnamefont{Hickel}},
  \bibinfo{author}{\bibfnamefont{J.}~\bibnamefont{Neugebauer}},
  \bibinfo{author}{\bibfnamefont{G.}~\bibnamefont{Kresse}},
  \bibinfo{author}{\bibfnamefont{A.}~\bibnamefont{Janotti}}, \bibnamefont{and}
  \bibinfo{author}{\bibfnamefont{C.~G.} \bibnamefont{{Van de Walle}}},
  \bibinfo{journal}{Rev. Mod. Phys.} \textbf{\bibinfo{volume}{86}},
  \bibinfo{pages}{253} (\bibinfo{year}{2014}),
  \urlprefix\url{https://link.aps.org/doi/10.1103/RevModPhys.86.253}.

\bibitem[{\citenamefont{Komsa et~al.}(2012)\citenamefont{Komsa, Rantala, and
  Pasquarello}}]{Komsa2012}
\bibinfo{author}{\bibfnamefont{H.~P.} \bibnamefont{Komsa}},
  \bibinfo{author}{\bibfnamefont{T.~T.} \bibnamefont{Rantala}},
  \bibnamefont{and}
  \bibinfo{author}{\bibfnamefont{A.}~\bibnamefont{Pasquarello}},
  \bibinfo{journal}{Phys. Rev. B} \textbf{\bibinfo{volume}{86}},
  \bibinfo{pages}{045112} (\bibinfo{year}{2012}), ISSN
  \bibinfo{issn}{10980121},
  \urlprefix\url{https://journals.aps.org/prb/abstract/10.1103/PhysRevB.86.045112}.

\bibitem[{\citenamefont{Wiktor et~al.}(2017)\citenamefont{Wiktor,
  Rothlisberger, and Pasquarello}}]{Wiktor2017}
\bibinfo{author}{\bibfnamefont{J.}~\bibnamefont{Wiktor}},
  \bibinfo{author}{\bibfnamefont{U.}~\bibnamefont{Rothlisberger}},
  \bibnamefont{and}
  \bibinfo{author}{\bibfnamefont{A.}~\bibnamefont{Pasquarello}},
  \bibinfo{journal}{J. Phys. Chem. Lett.} \textbf{\bibinfo{volume}{8}},
  \bibinfo{pages}{5507} (\bibinfo{year}{2017}),
  \urlprefix\url{https://pubs.acs.org/doi/abs/10.1021/acs.jpclett.7b02648}.

\bibitem[{\citenamefont{Li et~al.}(2017)\citenamefont{Li, Zhang, Zhang, Huang,
  Shen, Cheng, and Huang}}]{Li2017a}
\bibinfo{author}{\bibfnamefont{Y.}~\bibnamefont{Li}},
  \bibinfo{author}{\bibfnamefont{C.}~\bibnamefont{Zhang}},
  \bibinfo{author}{\bibfnamefont{X.}~\bibnamefont{Zhang}},
  \bibinfo{author}{\bibfnamefont{D.}~\bibnamefont{Huang}},
  \bibinfo{author}{\bibfnamefont{Q.}~\bibnamefont{Shen}},
  \bibinfo{author}{\bibfnamefont{Y.}~\bibnamefont{Cheng}}, \bibnamefont{and}
  \bibinfo{author}{\bibfnamefont{W.}~\bibnamefont{Huang}},
  \bibinfo{journal}{Appl. Phys. Lett.} \textbf{\bibinfo{volume}{111}},
  \bibinfo{pages}{162106} (\bibinfo{year}{2017}),
  \urlprefix\url{https://aip.scitation.org/doi/abs/10.1063/1.5001535}.

\bibitem[{\citenamefont{Xiao et~al.}(2019)\citenamefont{Xiao, Song, and
  Yan}}]{Xiao2019}
\bibinfo{author}{\bibfnamefont{Z.}~\bibnamefont{Xiao}},
  \bibinfo{author}{\bibfnamefont{Z.}~\bibnamefont{Song}}, \bibnamefont{and}
  \bibinfo{author}{\bibfnamefont{Y.}~\bibnamefont{Yan}}, \bibinfo{journal}{Adv.
  Mater.} \textbf{\bibinfo{volume}{31}}, \bibinfo{pages}{1803792}
  (\bibinfo{year}{2019}), ISSN \bibinfo{issn}{1521-4095},
  \urlprefix\url{https://onlinelibrary.wiley.com/doi/full/10.1002/adma.201803792}.

\bibitem[{\citenamefont{Wang et~al.}(2014)\citenamefont{Wang, Shao, Xie, Lyu,
  Liu, Gao, and Huang}}]{Wang2014}
\bibinfo{author}{\bibfnamefont{Q.}~\bibnamefont{Wang}},
  \bibinfo{author}{\bibfnamefont{Y.}~\bibnamefont{Shao}},
  \bibinfo{author}{\bibfnamefont{H.}~\bibnamefont{Xie}},
  \bibinfo{author}{\bibfnamefont{L.}~\bibnamefont{Lyu}},
  \bibinfo{author}{\bibfnamefont{X.}~\bibnamefont{Liu}},
  \bibinfo{author}{\bibfnamefont{Y.}~\bibnamefont{Gao}}, \bibnamefont{and}
  \bibinfo{author}{\bibfnamefont{J.}~\bibnamefont{Huang}},
  \bibinfo{journal}{Appl. Phys. Lett.} \textbf{\bibinfo{volume}{105}},
  \bibinfo{pages}{163508} (\bibinfo{year}{2014}), ISSN
  \bibinfo{issn}{00036951},
  \urlprefix\url{http://aip.scitation.org/doi/10.1063/1.4899051}.

\end{thebibliography}


\begin{thebibliography}{4}
\expandafter\ifx\csname natexlab\endcsname\relax\def\natexlab#1{#1}\fi
\expandafter\ifx\csname bibnamefont\endcsname\relax
  \def\bibnamefont#1{#1}\fi
\expandafter\ifx\csname bibfnamefont\endcsname\relax
  \def\bibfnamefont#1{#1}\fi
\expandafter\ifx\csname citenamefont\endcsname\relax
  \def\citenamefont#1{#1}\fi
\expandafter\ifx\csname url\endcsname\relax
  \def\url#1{\texttt{#1}}\fi
\expandafter\ifx\csname urlprefix\endcsname\relax\def\urlprefix{URL }\fi
\providecommand{\bibinfo}[2]{#2}
\providecommand{\eprint}[2][]{\url{#2}}

\bibitem[{\citenamefont{Walsh}(2017)}]{walsh_2017}
\bibinfo{author}{\bibfnamefont{A.}~\bibnamefont{Walsh}},
  \emph{\bibinfo{title}{{WMD-group/hybrid-perovskites/2015-fapi-PBEsol/FAPbI3-tetragonal-P42mnm.POSCAR.vasp}}}
  (\bibinfo{year}{2017}),
  \urlprefix\url{https://github.com/WMD-group/hybrid-perovskites/blob/master/2015_fapi_PBEsol/FAPbI3_tetragonal_P42mnm.POSCAR.vasp}.

\bibitem[{\citenamefont{Woodward}(1997{\natexlab{a}})}]{Woodward1997a}
\bibinfo{author}{\bibfnamefont{P.~M.} \bibnamefont{Woodward}},
  \bibinfo{journal}{Acta Crystallogr. Sect. B} \textbf{\bibinfo{volume}{53}},
  \bibinfo{pages}{32} (\bibinfo{year}{1997}{\natexlab{a}}),
  \urlprefix\url{https://doi.org/10.1107/S0108768196010713}.

\bibitem[{\citenamefont{Woodward}(1997{\natexlab{b}})}]{Woodward1997b}
\bibinfo{author}{\bibfnamefont{P.~M.} \bibnamefont{Woodward}},
  \bibinfo{journal}{Acta Crystallogr. Sect. B} \textbf{\bibinfo{volume}{53}},
  \bibinfo{pages}{44} (\bibinfo{year}{1997}{\natexlab{b}}),
  \urlprefix\url{https://doi.org/10.1107/S0108768196012050}.

\bibitem[{\citenamefont{Young and Rondinelli}(2016)}]{Young2016}
\bibinfo{author}{\bibfnamefont{J.}~\bibnamefont{Young}} \bibnamefont{and}
  \bibinfo{author}{\bibfnamefont{J.~M.} \bibnamefont{Rondinelli}},
  \bibinfo{journal}{J. Phys. Chem. Lett.} \textbf{\bibinfo{volume}{7}},
  \bibinfo{pages}{918} (\bibinfo{year}{2016}), ISSN \bibinfo{issn}{1948-7185},
  \urlprefix\url{https://pubs.acs.org/doi/10.1021/acs.jpclett.6b00094}.

\end{thebibliography}

\end{document}

\renewcommand{\arraystretch}{1.2}
\renewcommand{\thefigure}{S\arabic{figure}}
\renewcommand{\thetable}{S\arabic{table}}
\renewcommand{\theequation}{S\arabic{equation}}

\title{Supplemental material\\ The thermodynamic trends of intrinsic defects in primary halide perovskites: \\ A first-principles study}

\author{Haibo Xue}
\affiliation{Materials Simulation and Modelling, Department of Applied Physics, Eindhoven University of Technology, P.O. Box 513, 5600MB Eindhoven, the Netherlands.}
\affiliation{Center for Computational Energy Research, Department of Applied Physics, Eindhoven University of Technology, P.O. Box 513, 5600MB Eindhoven, the Netherlands.}

\author{Geert Brocks}
\affiliation{Materials Simulation and Modelling, Department of Applied Physics, Eindhoven University of Technology, P.O. Box 513, 5600MB Eindhoven, the Netherlands.}
\affiliation{Center for Computational Energy Research, Department of Applied Physics, Eindhoven University of Technology, P.O. Box 513, 5600MB Eindhoven, the Netherlands.}
\affiliation{Computational Materials Science, Faculty of Science and Technology and MESA+ Institute for Nanotechnology, University of Twente, P.O. Box 217, 7500AE Enschede, the Netherlands.}

\author{Shuxia Tao}
\email{s.x.tao@tue.nl}
\affiliation{Materials Simulation and Modelling, Department of Applied Physics, Eindhoven University of Technology, P.O. Box 513, 5600MB Eindhoven, the Netherlands.}
\affiliation{Center for Computational Energy Research, Department of Applied Physics, Eindhoven University of Technology, P.O. Box 513, 5600MB Eindhoven, the Netherlands.}

\maketitle

\clearpage

\tableofcontents

\clearpage

\section{The lowest energy structure of FAPbI$_3$}

\begin{figure} [h]
    \centering
    \includegraphics[width=0.8\textwidth]{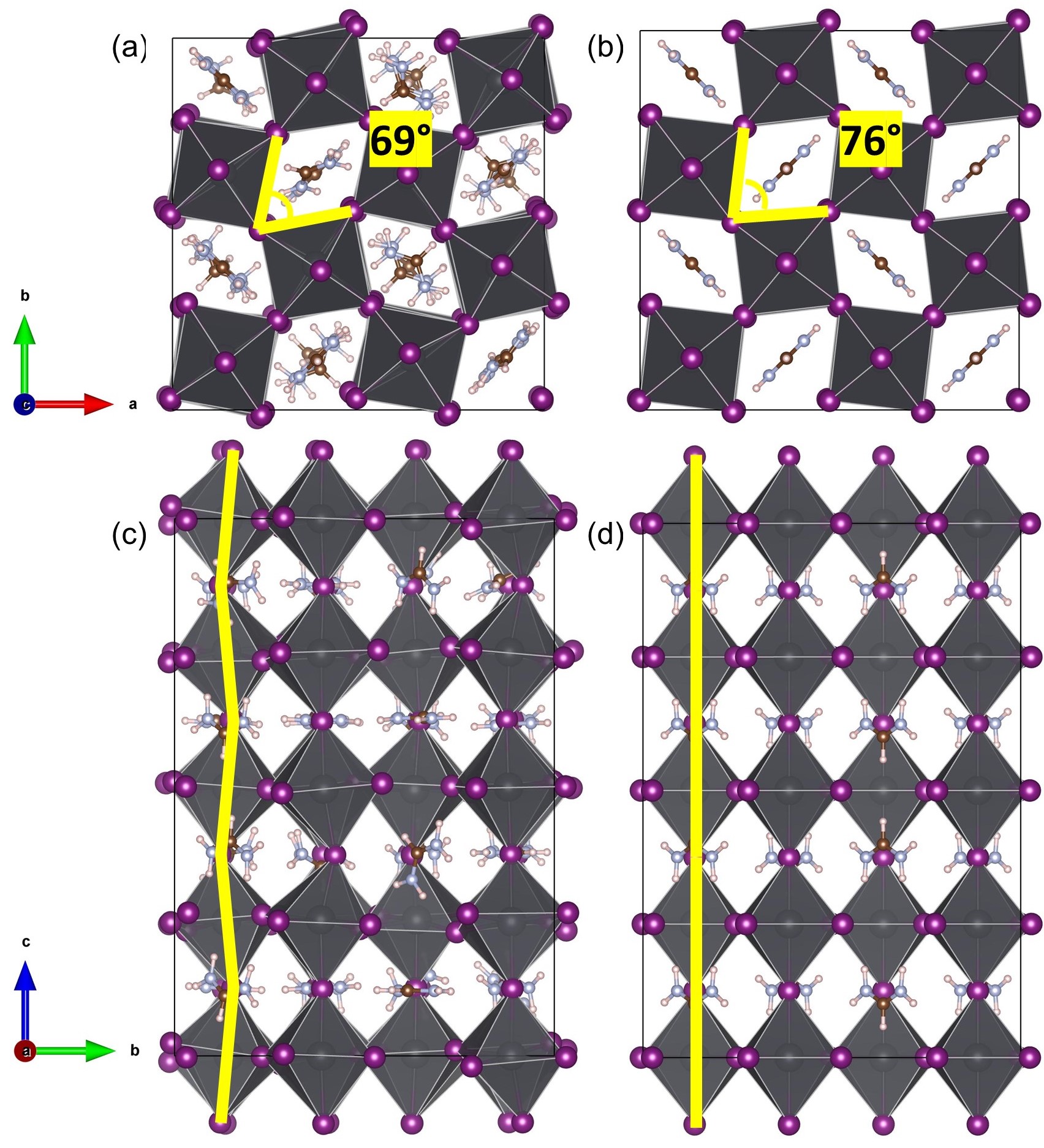}
    \caption{Comparison between FAPbI$_3$ structures (in a 2$\times$2$\times$2 tetragonal supercell); (a) and (c) are top and side views of our lowest energy structure of FAPbI$_3$. (b) and (d) are top and side views of the structure obtained from Ref.\cite{walsh_2017}.  The yellow lines in (a) and (b) illustrate the angles between the octahedras in the ab-plane, while in (c) and (d) they indicate the alignment of the octahedra with respect to the c-axis.}  \label{fig: Structures of FAPbI3}
\end{figure}

As discussed in the the main text, we create defects in FAPbI$_3$, starting from a $2\times 2\times 2$ tetragonal supercell built from the structure obtained from Ref. \onlinecite{walsh_2017}, which is shown in Fig. \ref{fig: Structures of FAPbI3} (b) and (d). We find that this structure indeed represents a local energy minimum. However, when optimizing the geometry upon introducing an interstitial, we find that the whole structure changes appreciably, accompanied a significant decrease of the energy. This shows that the starting structure used for FAPbI$_3$ is not the lowest energy structure. We then remove the interstitials again from the optimized interstitial-containing structures and relax them to obtain new bulk structures. It is found that these new bulk structures are all indeed more stable than the initial bulk structure. The lowest energy structure from this search is 0.3 eV per tetragonal unit cell lower in energy than the original structure.

The newly optimized structure is shown in Fig. \ref{fig: Structures of FAPbI3}(a) and (c). Comparing this structure to the original structure [Fig. \ref{fig: Structures of FAPbI3}(b) and (d)], one observes that in the $ab$-plane the octahedral tilting has increased, so that the smallest angle between the octahedra has decreased from 76$^\mathrm{o}$ to 69$^\mathrm{o}$. More striking is that the lower energy structure shows a wavy-pattern type of octahedral tilting along the c-axis [Fig. \ref{fig: Structures of FAPbI3}(c)], whereas in the higher energy structure the octahedra are all aligned along the c-axis [Fig. \ref{fig: Structures of FAPbI3}(d)]. Both of these distortion patterns are classical ways perovskite structures deal with ions of a non-ideal size \cite{Woodward1997a,Woodward1997b}. In this case the FA ion is too large to occupy the A cation position in the perovskite structure, giving a Goldschmidt tolerance factor $>$1, and the observed octahedral tilting pattern is typical for that case \cite{Woodward1997a,Woodward1997b,Young2016}. Although such deformations are also found for spherical A cations that are too big for the lattice \cite{Young2016}, we suggest that in this particular case they are also promoted by the (non-spherical) shape of the FA ion, and the ability of the latter to form hydrogen bonds with the surrounding iodide ions.

\clearpage

\section{Chemical potentials}

\begin{table*}[h]
    \caption{Chemical potentials of the M cations and halide anions of different perovskites at different growth conditions. The points A, B, and C refer to the points shown in Fig. 1 in the main text. The chemical potentials ($\Delta \mu$) of I, Br, Cl, Pb and Sn are referenced to the corresponding atoms in I$_2$, Br$_2$ and Cl$_2$ molecules and Pb and Sn metals, respectively.}
    \label{table: Chemical potentials}
    \begin{ruledtabular}
    \begin{tabular}{lcccccc}
        \multirow{2}{*}{Perovskite} & \multicolumn{2}{c}{Halide-rich (point A)} & \multicolumn{2}{c}{Halide-medium (point C)} & \multicolumn{2}{c}{Halide-poor (point B)}\\
          & $\Delta \mu _M$ (eV) & $\Delta \mu _X$ (eV) & $\Delta \mu _M$ (eV) & $\Delta \mu _X$ (eV) & $\Delta \mu _M$ (eV) & $\Delta \mu _X$ (eV)\\
        \hline
        MAPbI$_3$ & $-2.68$ & $0.00$ & $-1.34$ & $-0.67$ & $0.00$ & $-1.34$\\
        MAPbBr$_3$ & $-3.47$ & $0.00$ & $-1.74$ & $-0.87$ & $0.00$ & $-1.74$\\
        MAPbCl$_3$ & $-4.09$ & $0.00$ & $-2.05$ & $-1.02$ & $0.00$ & $-2.04$\\
        MASnI$_3$ & $-1.07$ & $-0.54$ & $-0.53$ & $-0.81$ & $0.00$ & $-1.08$\\
        FAPbI$_3$ & $-2.68$ & $0.00$ & $-1.34$ & $-0.67$ & $0.00$ & $-1.34$\\
        CsPbI$_3$ & $-1.71$ & $-0.49$ & $-0.85$ & $-0.91$ & $0.00$ & $-1.34$\\
    \end{tabular}
    \end{ruledtabular}
\end{table*}

\clearpage

\section{Optimized lattice parameters}

\begin{table}[h]
    \caption{Lattice constants of different perovskites optimized with the SCAN+rVV10 functional.}
    \label{table: Lattice constants}
    \begin{tabularx}{0.8\textwidth} { 
  >{\raggedright\arraybackslash}X 
  >{\centering\arraybackslash}X 
  >{\centering\arraybackslash}X}
        \hline
        \hline
        Perovskite & a=b (\AA) & c (\AA) \\
        \hline
        MAPbI$_3$ & 8.832 & 12.617  \\
        MAPbBr$_3$ & 8.381 & 11.974 \\
        MAPbCl$_3$ & 7.969 & 11.385 \\
        MASnI$_3$ & 8.783 & 12.548 \\
        FAPbI$_3$ & 8.936 & 12.637 \\
        CsPbI$_3$ & a=8.862, b=8.576 & 12.481 \\
        \hline
        \hline
    \end{tabularx}
\end{table}

\section{Defective structures}
\begin{figure}[h]
    \centering
    \includegraphics[width=0.8\textwidth]{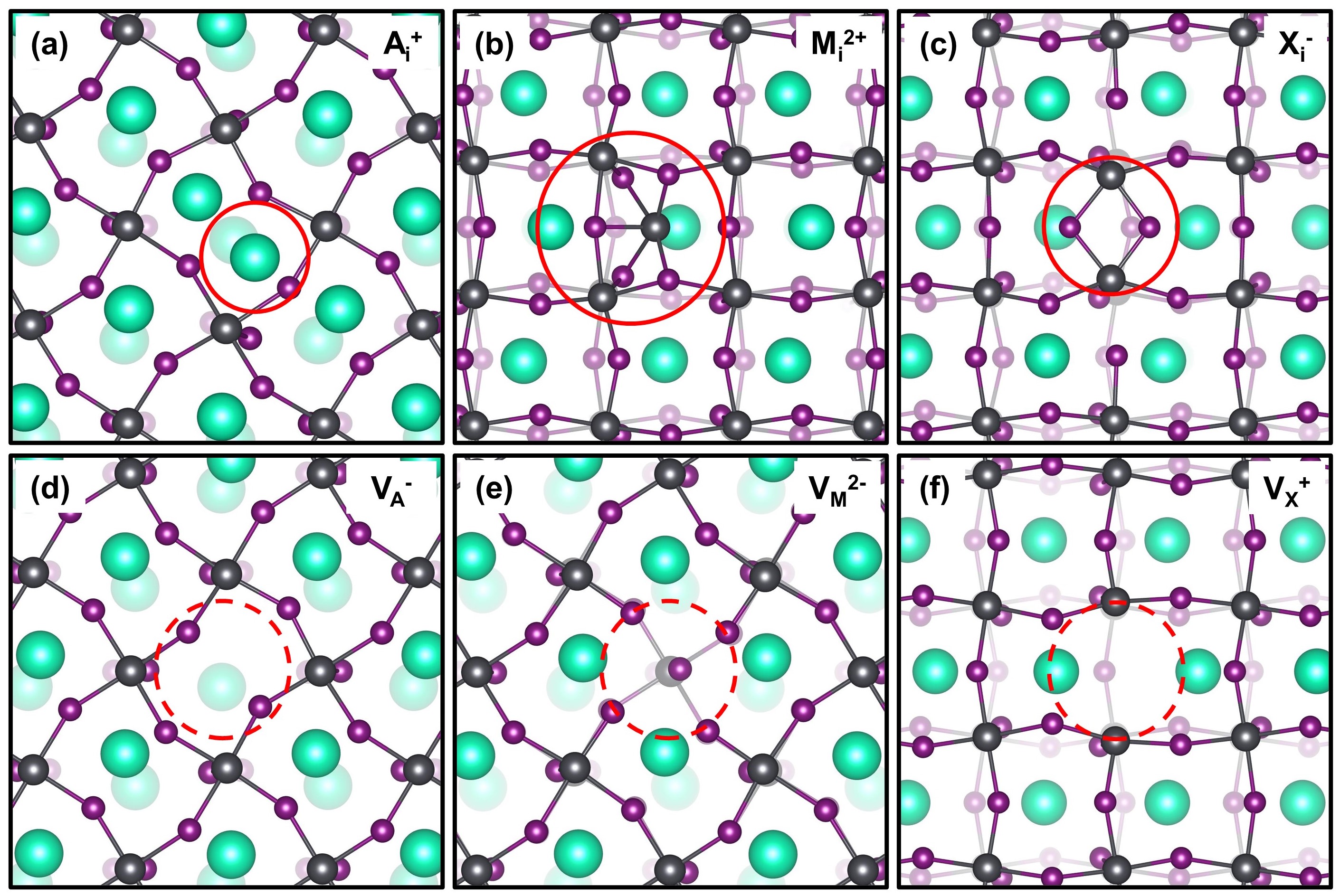}
    \caption{Optimized structures of different defects  in CsPbI$_3$ in their most stable charge states. From (a) to (f), they represent the interstitials A\textsubscript{i}$^+$, M\textsubscript{i}$^{2+}$, and X\textsubscript{i}$^-$, and the vacancies V\textsubscript{A}$^-$, V\textsubscript{M}$^{2-}$, and V\textsubscript{X}$^+$. The A cation, M cation and X anion interstitials are at the centers of the solid circles in (a), (b), and (c), respectively. The A cation, M cation and X anion vacancies are at the centers of the dashed circles in (d), (e), and (f), respectively.}
    \label{fig: Optimized structures of different defects}
\end{figure}

The same defects in the six perovskites we have studied generally have similar structures. Here, taking CsPbI$_3$ as an example, the optimized structures of different defects in their most stable charge states are visualized in Fig. \ref{fig: Optimized structures of different defects}. The interstitials typically occupy sites with less symmetry than the same ions in the original lattice [Fig. \ref{fig: Optimized structures of different defects}(a-c)].The A\textsubscript{i}$^+$ interstitial occupies the center of a line between two adjacent lattice-site A cations, and pushes the latter slightly aside [Fig. \ref{fig: Optimized structures of different defects}(a)]. The M\textsubscript{i}$^{2+}$ interstitial forms a square pyramidal bonding structures to five nearest neighbor X anions in the lattice [Fig. \ref{fig: Optimized structures of different defects}(b)]. The X\textsubscript{i}$^-$ interstitial forms a bridge-like structure with the adjacent lattice-site X anion, connecting two adjacent M cations, where the interstitial and the lattice site anion play an identical role [Fig. \ref{fig: Optimized structures of different defects}(b)] 

Vacancies usually only lead to mild changes of the local structures [Fig. \ref{fig: Optimized structures of different defects}(d-f)]. An A cation vacancy $\mathrm{V_A}^-$ only gives a small relaxation of the surrounding structure [Fig. \ref{fig: Optimized structures of different defects}(d)]. A M cation vacancy V\textsubscript{M}$^{2-}$ gives a more or less uniform contraction of the six adjacent M-X bonds [Fig. \ref{fig: Optimized structures of different defects}(e)], and a X vacancy V\textsubscript{X}$^+$ pushes the previously bonded M cations slightly away [Fig. \ref{fig: Optimized structures of different defects}(f)].

\clearpage

\section{Defect formation energies at different chemical potentials}

\begin{figure}[h]
    \centering
    \includegraphics[width=0.68\textwidth]{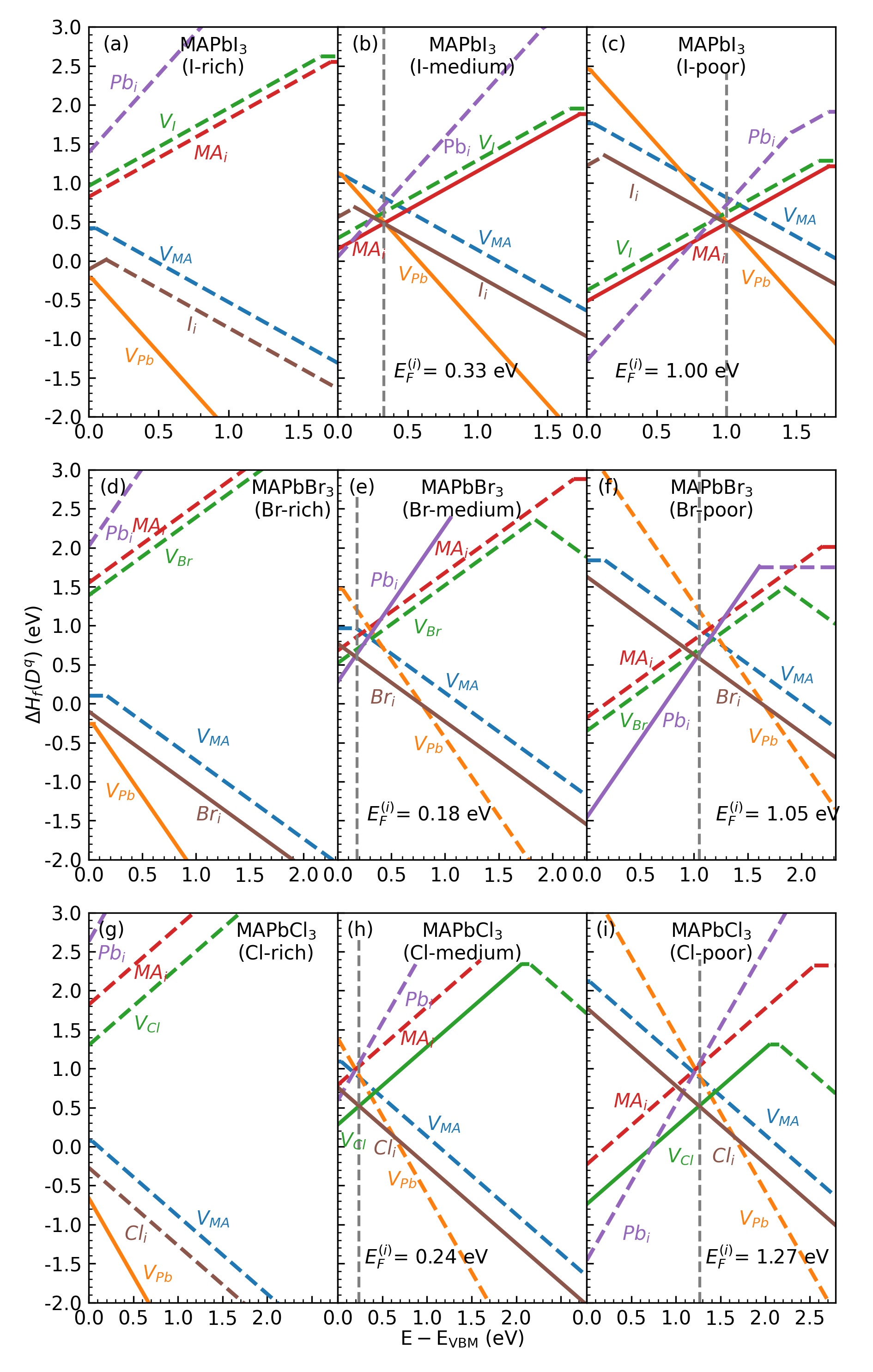}
    \caption{Defect formation energies of MAPbX$_3$ calculated at halide-rich, halide-medium and halide-poor growth conditions.}
    \label{fig: DFE of MAPbX3 in different grwoth conditions}
\end{figure}

\clearpage

\begin{figure}
    \centering
    \includegraphics[width=0.68\textwidth]{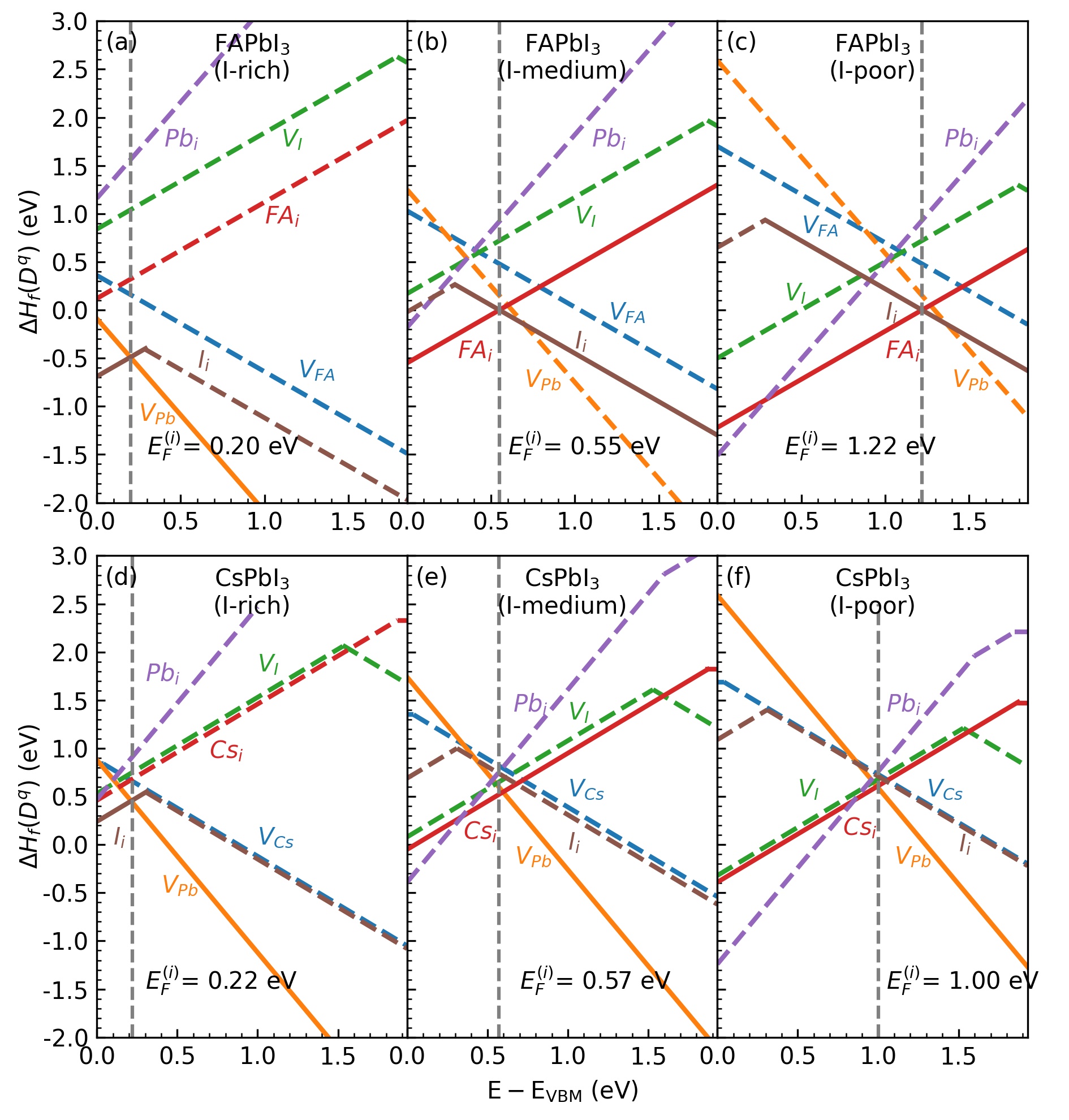}
    \caption{Defect formation energies of FAPbI$_3$ and CsPbI$_3$ calculated at halide-rich, halide-medium and halide-poor growth conditions.}
    \label{fig: DFE of MASnI3 in different grwoth conditions}
\end{figure}

\begin{figure}
    \centering
    \includegraphics[width=0.68\textwidth]{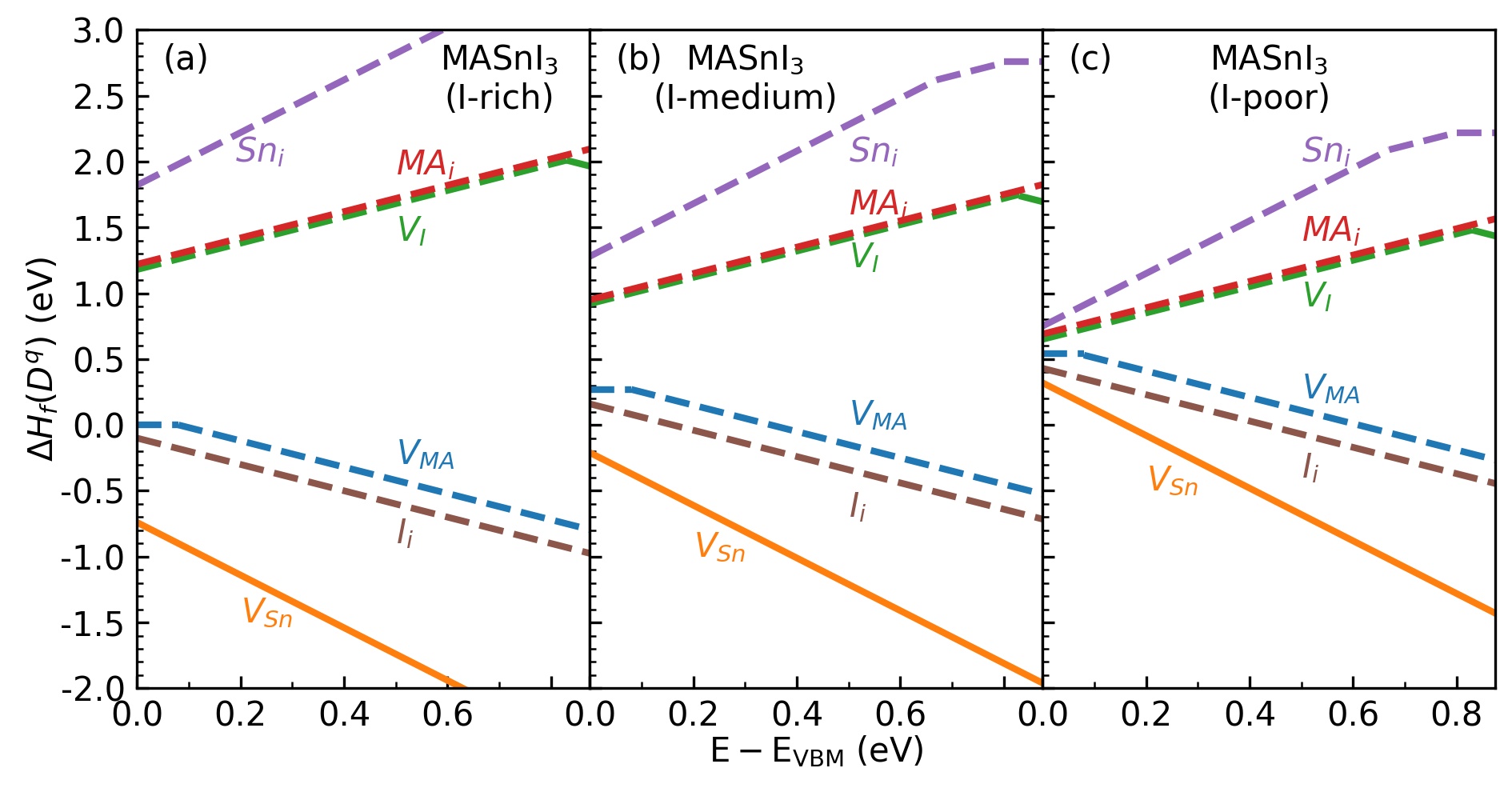}
    \caption{Defect formation energies of MASnI$_3$ calculated at halide-rich, halide-medium and halide-poor growth conditions.}
    \label{fig: DFE of MASnI3 in different grwoth conditions}
\end{figure}

\clearpage

\section{Neutral defects}

\begin{table}[h]
    \caption{Formation energies $\Delta H_f$ (eV) and concentrations $c$ (cm$^{-3}$) at $T$ = 300 K  of neutral defects in six perovskites calculated using the SCAN+rVV10 functional. The results of MASnI$_3$ are calculated based upon I-poor conditions, while those of the other perovskites are calculated based on halide-medium conditions.}
    \label{table: DFE_neutral }
    \begin{ruledtabular}
    \begin{tabular}{lllllll}
        Perovskites & $\mathrm{V_{A}}^0$ & $\mathrm{V_{M}}^0$ & $\mathrm{V_{X}}^0$ & $\mathrm{A_i}^0$ & $\mathrm{M_i}^0$ & $\mathrm{X_i}^0$\\
        \hline
    \multicolumn{7}{l}{Defect formation energy $\Delta H_f$ (eV)}\\
    MAPbI$_3$ & 1.09 & 1.15 & 1.95 & 1.88 & 3.25 & 1.00 \\
    MAPbBr$_3$ & 0.97 & 1.47 & 2.48 & 2.88 & 3.49 & 0.77 \\
    MAPbCl$_3$ & 1.09 & 1.59 & 2.34 & 3.34 & 4.17 & 0.77 \\
    MASnI$_3$ & 0.54 & 0.30 & 1.47 & 1.57 & 2.22 & 0.43 \\
    FAPbI$_3$ & 1.05 & 1.21 & 1.96 & 1.41 & 3.15 & 0.75 \\
    CsPbI$_3$ & 1.26 & 1.71 & 1.79 & 1.90 & 3.07 & 1.29 \\
    \multicolumn{7}{l}{Defect concentration $c$ (cm$^{-3}$)}\\
    MAPbI$_3$ & 2.11$\times 10^{3}$ & 1.90$\times 10^{2}$ & 2.45$\times 10^{-11}$ & 3.26$\times 10^{-10}$ & 3.06$\times 10^{-33}$ & 1.80$\times 10^{5}$ \\
    MAPbBr$_3$ & 2.31$\times 10^{5}$ & 8.84$\times 10^{-4}$ & 2.75$\times 10^{-20}$ & 5.57$\times 10^{-27}$ & 3.34$\times 10^{-37}$ & 1.39$\times 10^{9}$ \\
    MAPbCl$_3$ & 2.33$\times 10^{3}$ & 1.15$\times 10^{-5}$ & 9.16$\times 10^{-18}$ & 1.11$\times 10^{-34}$ & 1.28$\times 10^{-48}$ & 1.91$\times 10^{9}$ \\
    MASnI$_3$ & 3.95$\times 10^{12}$ & 3.49$\times 10^{16}$ & 2.94$\times 10^{-3}$ & 5.45$\times 10^{-5}$ & 5.25$\times 10^{-16}$ & 7.75$\times 10^{14}$ \\
    FAPbI$_3$ & 7.59$\times 10^{3}$ & 2.24$\times 10^{1}$ & 1.50$\times 10^{-11}$ & 2.55$\times 10^{-2}$ & 1.27$\times 10^{-31}$ & 3.41$\times 10^{9}$ \\
    CsPbI$_3$ & 2.38$\times 10^{0}$ & 7.03$\times 10^{-8}$ & 1.04$\times 10^{-8}$ & 1.48$\times 10^{-10}$ & 3.92$\times 10^{-30}$ & 2.36$\times 10^{0}$ \\
    \end{tabular}
    \end{ruledtabular}
\end{table}

\clearpage

\bibliography{SI}